\documentclass[preprint,sort&compress]{elsarticle}
\usepackage{amssymb,amsthm,amsmath,commath,bbm,psfrag,setspace,color}
\usepackage{array,booktabs,multirow,url,float,epstopdf}
\usepackage[FIGTOPCAP,nooneline]{subfigure}
\usepackage{color}
\usepackage[pdftex,colorlinks]{hyperref}
\definecolor{darkblue}{cmyk}{1,0,0,0.8}
\definecolor{darkred}{cmyk}{0,1,0,0.7}
\hypersetup{anchorcolor=black,
  citecolor=darkblue, filecolor=darkblue,
  menucolor=darkblue,pagecolor=darkblue,urlcolor=darkblue,linkcolor=darkblue}
\usepackage{cite}

\usepackage{amsmath}
\DeclareMathOperator{\sign}{sign}
\usepackage{algorithm}
\usepackage{algorithmic}
\usepackage{xcolor}

\DeclareMathOperator{\bigo}{\mathcal{O}}
\DeclareMathOperator{\re}{\mathbbm{R}}
\DeclareMathOperator{\nat}{\mathbbm{N}}

\newcommand{\js}[1]{{\textcolor{black}{#1}}}
\newcommand{\jsr}[2]{{\textcolor{black}{#1}}}

\theoremstyle{definition}

\begin{document}

\begin{frontmatter}


\title{Numerical analysis of a multistable capsule system under the delayed feedback control with a constant delay}

\author[EXT]{Zhi Zhang}
\ead{zz326@exeter.ac.uk}

\author[ESPOL,CFD]{Joseph P\'aez Ch\'avez}
\ead{jpaez@espol.edu.ec}

\author[EXT]{Jan Sieber}
\ead{j.sieber@exeter.ac.uk}

\author[EXT]{Yang Liu \corref{mycorrespondingauthor}}
\ead{y.liu2@exeter.ac.uk}

\address[EXT]{Faculty of Environment, Science and Economy, University of Exeter, North Park Road, Exeter EX4 4QF, UK}

\address[ESPOL]{Center for Applied Dynamical Systems and Computational Methods (CADSCOM), Faculty of Natural Sciences and Mathematics, Escuela Superior Polit\'ecnica
del Litoral, P.O. Box 09-01-5863, Guayaquil, Ecuador}
\address[CFD]{Center for Dynamics, Department of Mathematics, TU Dresden, D-01062 Dresden, Germany}

\cortext[mycorrespondingauthor]{Corresponding author. Tel:
+44(0)1392-724654, e-mail: y.liu2@exeter.ac.uk.}

\begin{abstract}
The vibro-impact capsule system is a self-propelled mechanism that
has abundant coexisting attractors and moves rectilinearly under
periodic excitation when overcoming environmental resistance. In this paper, we study the control of coexisting attractors in this system by using a delayed
feedback controller (DFC) with a constant delay. \js{The aim of our control is to steer}
this complex system \jsr{toward}{to present} \jsr{an attractor with preferable performance characteristics}{a preferred performance} among
multiple coexisting attractors, e.g., a periodically fast forward progression. For this purpose, we \jsr{give an example of a feedback-controlled transition}{exemplify the control} from a period-3 motion with low progression speed to a period-1 motion with high progression speed
at \js{the system parameters} where both responses coexist. The effectiveness of this
controller is investigated numerically by considering its
convergence time and the required \js{control} energy \js{input} to \jsr{achieve transition}{complete this task}.
\jsr{We combine pseudo-spectral approximation of the delay, event detection for the discontinuities and path-following}{Also, the path-following} (continuation) techniques for
non-smooth delay dynamical systems to carry out bifurcation analysis. \jsr{We}{to} systematically study the dynamical performance of the controlled system \jsr{when}{by} varying its control gain and
delay time. Our numerical simulations show the effectiveness of DFC under a wide range of system parameters. \jsr{We find}{It is found} that \jsr{the}{our} desired period-1 motion is achievable \jsr{in}{at} a \jsr{}{specific} range of \jsr{}{the} control \jsr{delays}{delay}
between \jsr{a}{the occurrences of a} period-doubling and a grazing
bifurcation\jsr{}{s}. Therefore, two-parameter continuation of these two
bifurcations with respect to the control delay and control gain is
conducted \jsr{to identify}{, and} the delay-gain parameter region \jsr{where the period-1 motion is stable}{is identified}. The
findings of this work can be used for tuning control parameters
in experiments, and similar \jsr{analysis}{discussion and analysis} can be carried out \jsr{for other}{to study other similar} non-smooth dynamical
systems with a constant delay term.
\end{abstract}


\end{frontmatter}

\section{Introduction}
The vibro-impact capsule system \cite{liu2013modelling, nguyen2008experimental} is a typical piecewise-smooth dynamical system, which consists of a periodically excited inner
mass interacting with the main body in the presence of environmental
resistance for rectilinear progression. The vibro-impacting driven
concept has been adopted to design the self-propelled capsule system
for active gastrointestinal endoscopy
\cite{guo2020experimental}. However, due to its high
degree of nonlinearity, the dynamics of the system may witness a
significant change when any small variation from the environment or
in the system parameter is encountered. For example, in
\cite{liu2015forward}, the system has two
coexisting attractors near a grazing event with fractal basins of
attraction, indicating that a small perturbation on its initial
conditions may result in different dynamical responses. In
\cite{chavez2016path}, by changing the frequency or the amplitude of
excitation slightly, the direction of capsule's drift can be
switched, if the stiffness ratio of the system is kept at a low
value. Because of its complex dynamics and high sensitivity to
parameter variation, the dynamics of such vibration-driven capsule
systems have received extensive attention from many researchers in
the past decade.
For example, in \cite{chernous2002optimum}, Chernousko presented that a two-mass system can be driven by the internal force from interaction between two main bodies, which becomes the basic concept for developing self-driven capsule system.
Following this concept, Li \emph{et al}. built a prototype capsule robot driven by a magnet rod and proposed a strategy to achieve the one-dimensional motion \cite{li2006motion}.
In \cite{gu2018dynamical}, Gu and Deng found that the random stick-slip phenomena appears in the capsule system with Hertzian contact and random environmental perturbation, and stochastic P-bifurcation may happen by varying the system parameters.
On the other hand, Zhao and Ouyang \cite{zhao2021capsule} theoretically studied a capsule-structured triboelectric energy harvester and showed that its performance was not limited by its frequency bandwidth, which was different from other types of vibration-based energy harvesters.
For these vibration-driven systems, multistability is an intrinsic property
associating with the possession of two or more coexisting attractors
for a given set of parameters.
For example, in \cite{zhang2018multistability}, a three-degree-of-freedom vibro-impact system coexists tori $T^3$ attractor and other types of attractors, such as period-3 and  tori $T^2$ near the codimension-3 bifurcation.
In \cite{xu2019analytical}, Xu and Ji showed that a vibro-impact system coexists stable, and unstable impact motions and the non-impact periodic motion
were observed in a small range between the grazing and the saddle-node bifurcations.
As a consequence, the capsule system
can present different performances based on different attractors.
For example, some attractor presents forward or backward
progression, while some performs periodic or chaotic motion, under
the same values of system parameters \cite{liu2015forward}. In
addition, the coexisting attractors may have different motion
efficiencies in terms of energy consumption and progression rate,
even that they all present periodic responses with forward
progressions. Hence, from a practical point of view, controlling
multistability is useful to determine the motion of the capsule
system and to improve its working efficiency, in particular for the
self-propelled capsule endoscopy in the intestinal environment. To
this end, developing a reliable control method for the vibro-impact
capsule system is crucial.

From an engineering perspective, there are two major issues related
to multistability. On the one hand, the performance of multistable
systems can be easily altered without changing their control
parameters, e.g.,
\cite{pavlovskaia2015modelling,Serdukova2021}. On the other
hand,  some of the coexisting attractors of these systems may
correspond to the states causing costly failure.
In details, for the application of Jeffcott rotor system, at some cases with bistability,
grazing-induced chaotic motion treated as undesired motion, and dangerous vibration should be avoided
\cite{mora2020}. Hence, it is necessary to control the system to avoid this type of parameter region and ensure systems have high efficient performance.
Similar problems can be found in drill-strings system, where the coexisting stick-slip oscillations causing bit wear and inefficient drilling should be avoided \cite{liu2017}.
Thus, control of multistability is essential for maintaining system
stability while improving its performance. For this reason, it has
received considerable attention from the nonlinear dynamics research
community in the past few decades, see
\cite{makarenkov12,pisarchik2014control}. Many useful control
methods were developed to control the multistability of dynamical
system for presenting a desired motion. For example, in
\cite{ott1990controlling}, the OGY method was proposed to stabilise
chaotic attractors to an embedded unstable periodic attractor, which
was adopted later for controlling chaos in different dynamical
systems \cite{lai2011transient}.
In terms of non-smooth dynamical systems, the OGY method presented good performance in many applications, such as the mass-beam system with an elastic stop \cite{haiyan1995controlling}, the piecewise linear oscillator \cite{kleczka1992local,de2004controlling} and the impact-friction oscillator \cite{begley2001ogy}.
Besides that, in \cite{pisarchik2014control}, Pisarchik and Feudel presented an
efficient way to control the multistability based on annihilating
all undesired coexisting attractors. However, this could
result in the alterations of existing structure of dynamical
systems, thus the basins of attraction, indicating the
original dynamics cannot be preserved by introducing this method.
Similar features and control results can be found in bifurcation control methods, see e.g., \cite{ji2002bifurcation,ji2001local}, which focus on controlling the appearance of some bifurcations, such as the pitchfork and the saddle-node bifurcations, to enlarge the stable region of the desired motion or postpone the appearance of the jump and hysteresis phenomena.
In order to avoid this problem, some new control methods were
developed to not only achieve the control of multistability, but
also keep the original dynamics of the controlled system. In
\cite{liu2017controlling}, a linear gain feedback control based on
the feedback of the desired attractor was used to control the
coexisting attractors in an impacting system. In
\cite{zhang2022controlling}, a parameter control method was
developed to generate a continuous path between the coexisting
attractors for a class of non-autonomous dynamical systems. Wang
\emph{et al}. \cite{wang2016geometrical} applied a parameter
perturbation method to the nonlinear dynamical networks and achieved
the switching from one attractor to the desired one. For a similar
network, Cornelius \emph{et al}. \cite{cornelius2013realistic}
developed a feedback control method which generates a small
perturbation on system state to achieve the switching among
coexisting attractors. In \cite{zhang2021controlling}, a delay
feedback control was applied to a soft impact oscillator to control
its coexisting attractors near grazing.

\paragraph{\js{Delayed feedback control (DFC)}}

\js{The present work will consider} a DFC with Pyragas-type delay feedback control for controlling multistability in a vibro-impact capsule system. \js{DFC is of the form} \cite{pyragas1992continuous}
\js{\begin{displaymath}
  u(t)=K\left(y(t)-y(t-\tau_\mathrm{d})\right),
\end{displaymath}
where $K$, $\tau_\mathrm{d}$ and $y$ are suitably chosen control gain, delay and output of the system, respectively.} \jsr{Vibro-impact capsules}{based on the similar concept will be considered}
\jsr{pose}{which has} a number of \jsr{}{control} issues \jsr{potentially addressable by feedback control}{in this capsule system}, such as choosing \jsr{between}{the} forward or backward motion \cite{liu2015forward} and the control of multistability \cite{liu2017b}.
These applications are meaningful, since this control principle can be the potential tool to control the multistability of many different systems based on the concept of vibro-impact capsule system, such as the worm-like robots \cite{xue2022coordinated}, the deformable robots with a magnetizable material \cite{tkachenko2022mathematical}, and the two-sided vibro-impact energy harvester \cite{zhao2021capsule,dulin2022improving}.
In addition, there are many advantages to applying the DFC for the control of multistability.
For example, DFC can make the controlled
non-smooth dynamical systems monostable, even at the
near-grazing regime, \jsr{simplifying}{indicating that it simplifies} the complex
dynamics while keeping the controlled system stable at the
desired motion. Also, compared with the classical PID control method \cite{gupta2020nonlinear} and the parameter control method \cite{zhang2022controlling}, this method does not need to have the detailed
information of the desired attractor and trajectory as the reference signal, but only requires the information of
desired period of targeting attractor \js{(set equal to the delay \js{$\tau_\mathrm{d}$} in the control)}.
DFC is easy to \jsr{be implemented for}{be introduced to the} complex nonlinear dynamical systems\jsr{,}{ on controlling multistability}\jsr{ because it only}{ and just} needs partial information of system states \cite{Sudharsan2022,Zheng2022}, compared with \js{the} LQR method which \jsr{designs}{focuses on optimising} the controller for linear systems to \jsr{minimise a}{satisfy the} cost function \cite{naidu2002optimal}.
\jsr{U}{Besides that, u}nlike traditional vibration control strategies, such as passive \cite{franchek1996adaptive}, active and semi-active control \cite{liu2005comparison}, which \jsr{introduce}{are emphasis on introducing the} significant changes to system's structure and keep the controlled system away from its original dynamics\jsr{, a system  with}{. By contrast, the} DFC \jsr{can asymptotically converge to the}{does not need to change the controlled system's basic structures, and the controlled system with the DFC can still remain some} desired dynamics of the original \js{uncontrolled} system.
Thus, both benefits motivate us to use the DFC to control the capsule system.

\jsr{}{It is well-known that the} DFC was first
introduced by Pyragas \cite{pyragas1992continuous} \jsr{to}{, and can be used
to} stabilise chaos to an (originally) unstable periodic orbit. When
introducing this controller into the vibro-impact capsule system,
the finite dimensional non-smooth dynamical system becomes an infinite-dimensional system. Undoubtedly, the combination of this change and
the non-smooth properties may induce new difficulties for analysing
the time-delayed non-smooth dynamical system
\cite{zhang2020calculating}. These difficulties may lead the analysis of non-smooth delay differential equations (DDEs) to failure through utilising the existing numerical continuation software packages directly (e.g., DDE-BIFTOOL
\cite{engelborghs2002numerical,sieber2017dde}, PDDE-CONT
\cite{szalai2006continuation} and Knut \cite{szalai2013dde}).
Therefore, the path-following analysis based on the approximate
numerical approach proposed in \cite{joseph2020} will be employed to
study the dynamics of the vibro-impact capsule system under the
time-delayed feedback control. In addition, in the present work,
we will study the effectiveness of the proposed controller with the consideration of some specific performance indices, so optimising both the capsule system and the controller.


The main contribution of the present paper is to employ the
DFC with a constant delay to control the motion of the
vibro-impact capsule system via controlling its coexisting
attractors. To this end, we will adopt a numerical approach based on
path-following techniques to analyse the dynamical responses of the capsule system. The rest of the paper is organised as follows. In Section 2, the physical model and equations of motion of the vibro-impact capsule system are introduced, as well as the mathematical formulation of the DFC. In Section 3, bifurcations of the vibro-impact capsule system with power spectrum density are analysed numerically, and the dynamical performance of the controlled system are studied numerically with respect to some specific performance indices, such as the convergence time and the control energy. In Section 4, further numerical bifurcation analysis of the controlled system is carried out by using the path-following (continuation) methods. Finally, conclusions of the present work are given in Section 5.


\section{Mathematical model}

\begin{figure}[h!]
\centering
\includegraphics[width=0.7\textwidth]{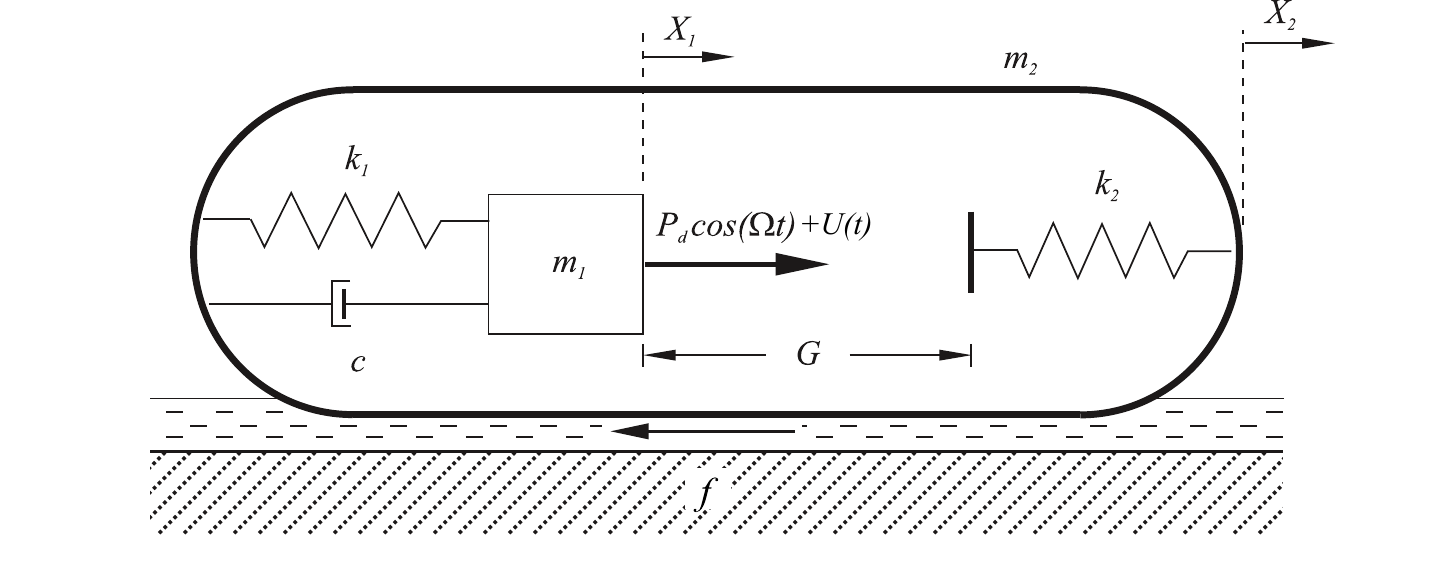}
\caption{{\color{black}Physical model of the vibro-impact capsule
system.}}\label{fig_capsule}
\end{figure}

In this paper, a vibro-impact capsule system depicted in
Fig.~\ref{fig_capsule} is considered. This system consists of
two parts: an inner actuator and a rigid capsule. In details,
the main part of the inner actuator is the internal mass
$m_1$ driven by an external harmonic force with amplitude $P_d$ and
frequency $\Omega$, and a weightless plate. This inner mass is connected to the rigid capsule via a linear spring with stiffness $k_1$ and a viscous damper with damping
coefficient $c$, and the plate is connected with the capsule via a secondary linear spring with stiffness $k_{2}$. The absolute displacements of the internal mass and the
capsule denote by $X_1$ and $X_2$, respectively. If
$X_{1}-X_{2}\ge G$, where $G$ is the gap, the weightless plate can
be hit by the internal mass $m_1$. The capsule starts to move
forward or backward until the elastic force acting on the capsule exceeds
the threshold of the dry friction force $P_f$ between the capsule
and supporting surface. In this study, we use Coulomb friction to calculate the frictional force between the capsule and the supporting surface as
\begin{equation}\label{columb}
f=-\sign (\dot{X}_2)\cdot P_{f},~\dot{X}_{2}\ne 0,
\end{equation}
and $f$ equals to the net elastic force from the inner mass and is with the opposite direction of this force, if $\dot{X}_{2}=0$,
where $P_f=\mu (m_1+m_2)g$, $\mu$ is the friction coefficient
between the capsule and the sliding surface, and $g$ is the gravitational acceleration.
Thus, the dimensional equation of the motion with {\color{black}a controller $U(t)$} can be constructed as {\color{black}as follows \cite{liu2013modelling}.
\begin{itemize}
\item[(i)] When the contact between the mass $m_1$ and the plate dose not happen ($X_{1}-X_{2}<G$) and the capsule $m_{2}$ is stationary,
the motion of the internal mass $m_{1}$ and the capsule $m_{2}$ can be expressed as
\begin{align}\label{nocontact}
m_{1}\ddot{X}_{1}=&\, P_{d} \cos(\Omega t)+U(t)+k_{1}(X_{2}-X_{1})+c(\dot{X}_{2}-\dot{X}_{1}),\\
\dot{X}_{2}=&\, 0,
\end{align}
where the threshold of the dry friction force is larger than the net elastic force acting on the capsule, i.e., $|k_{1}(X_{2}-X_{1})+c(\dot{X}_{2}-\dot{X}_{1})|<P_{f}$.

\item[(ii)] When there is no contact between the mass $m_1$ and the plate ($X_1-X_2<G$) and the capsule is moving, the equation of motion of the internal mass can be expressed as Eq.~\eqref{nocontact}, and the capsule can be written as
\begin{align}\label{nocontact_move}
&m_{1}\ddot{X}_{2}= -\sign(\dot{X}_{2})\cdot P_{f} -k_{1}(X_{2}-X_{1})-c(\dot{X}_{2}-\dot{X}_{1}).
\end{align}

It is worth noting that
the dry friction force acting on the capsule has an opposite direction with the net elastic force acting on the capsule, when the net elastic force just equals to the threshold of the dry friction force, i.e., $|k_{1}(X_{2}-X_{1})+c(\dot{X}_{2}-\dot{X}_{1})|= P_{f}$, and the capsule begins to move. At this moment, the dry friction force can be described as
\begin{align}\label{nocontact_dryfriction}
f=-\sign(k_{1}(X_{2}-X_{1})+c(\dot{X}_{2}-\dot{X}_{1}))\cdot P_{f}.
\end{align}

\item[(iii)]  When the contact between the mass $m_1$ and the plate occurs ($X_1-X_2\ge G$) and the threshold of the dry friction force is larger than
the net elastic force acting on the capsule ($|k_{1}(X_{2}-X_{1})+c(\dot{X}_{2}-\dot{X}_{1})-k_{2}(X_{1}-X_{2}-G)|<P_{f}$), the capsule is stationary, and the equations of motion of the internal mass and the capsule can be written as
\begin{align}\label{contact}
m_{1}\ddot{X}_{1}=&\, P_{d} \cos(\Omega t)+U(t)+k_{1}(X_{2}-X_{1})+c(\dot{X}_{2}-\dot{X}_{1})-k_{2}(X_{1}-X_{2}-G),\\
\dot{X}_{2}=&\, 0.
\end{align}

\item[(iv)] When there is a contact between the mass $m_1$ and the plate ($X_1-X_2\ge G$) and the capsule is moving, the equation of motion of the internal mass can be written as Eq.~\eqref{contact}, and the capsule can be expressed as
\begin{align}\label{contact_move}
&m_{1}\ddot{X}_{2}= -\sign(\dot{X}_{2})\cdot P_{f} -k_{1}(X_{2}-X_{1})-c(\dot{X}_{2}-\dot{X}_{1})+k_{2}(X_{1}-X_{2}-G).
\end{align}
Similarly as phase (ii), the direction of dry friction force acting on the capsule should be opposite to the net elastic force acting on the capsule, when the net elastic force just equals to the threshold of the dry friction force ($|k_{1}(X_{2}-X_{1})+c(\dot{X}_{2}-\dot{X}_{1})-k_{2}(X_{1}-X_{2}-G)|= P_{f}$), and the capsule begins to move. At this moment, the dry friction force can be written as
\begin{align}\label{contact_dryfriction}
f=-\sign(k_{1}(X_{2}-X_{1})+c(\dot{X}_{2}-\dot{X}_{1})-k_{2}(X_{1}-X_{2}-G))\cdot P_{f}.
\end{align}
\end{itemize}
}

To simplify the analysis, the {\color{black} above}
equations of motion will be transformed into dimensionless form
based on the following nondimensional variables
\begin{align*}
\tau&=\Omega_0 t,&x_i&=\frac{k_{1}}{P_{f}}X_i,& y_{i}&=\frac{d
x_{i}}{d \tau} = \frac{k_{1}}{\Omega_0 P_{f}}\dot{X}_{i},&
\dot{y}_{i}&=\frac{d y_{i}}{d \tau}=\frac{k_{1}}{\Omega^{2}_{0}
  P_{f}} \ddot{X}_{i},
\end{align*}
(for $i=1,2$) and parameters
\begin{align*}
 \Omega_0&=\sqrt{\frac{k_{1}}{m_{1}}},&\omega&=\frac{\Omega}{\Omega_{0}},&\alpha&=
 \frac{P_d}{P_{f}},& \delta&=\frac{k_{1}}{P_{f}}G,&\beta&=\frac{k_{2}}{k_{1}},&\zeta&=
 \frac{c}{2 m \Omega_0},& \gamma&=\frac{m_{2}}{m_{1}}.
\end{align*}
The resulting system variables and parameters are summarised in Table~\ref{tab:syspars}.
\begin{table}[ht]
  \centering
  \caption{Non-dimensionalised time, state variables and system parameters.}
  \vspace{-5pt}
  \small
  \setstretch{1.2}
  \begin{tabular}[t]{c|c|l}
    \hline
    \textbf{Variable}&\textbf{Value}&\textbf{Description}\\ \hline
    $\tau$&$\mathbb{R}$&Time in multiples of natural period $1/\Omega_0$ of internal mass-spring
    system$\phantom{\int^1}$\\
    $x_1$, $y_1$&$\mathbb{R}$ & Displacement, velocity of internal mass\\
    $x_2$, $y_2$&$\mathbb{R}$ & Displacement, velocity of capsule shell\\[0.5ex]
    $\Omega_0$&&Natural frequency of internal mass-spring system\\
    $\omega$&$[0.93,{\color{black}1.03}]$& Forcing frequency in multiples of natural frequency $\Omega_0$\\
    $\alpha$&$1.6$&Forcing amplitude relative in multiples of static friction force $P_\mathrm{f}$\\
    $\delta$&$0.02$&Gap between plate and internal mass at rest\\
    $\beta$&$15$& Spring stiffness of plate relative to stiffness of internal mass-spring system\\
    $\zeta$&$0.01$& Relative damping of internal mass-spring system \\
    $\gamma$&$5$& Weight of the rigid capsule relative to the weight of internal mass \\\hline
  \end{tabular}
  \label{tab:syspars}
\end{table}

The operation of the vibro-impact capsule system with {\color{black}the DFC} {\color{black}in different phases can be expressed as below.}
\begin{itemize}
\item[(i)] \emph{No contact with stationary capsule}: $x_2-x_1<\delta$, $y_2=0$,
\begin{align}\label{nomovingcapsu_nocontactdelay}
&\dot{x}_{1}=y_{1},\nonumber\\
&\dot{y}_{1}=\alpha \cos(\omega \tau)+(x_{2}-x_{1})+2\zeta(y_{2}-y_{1})+u(\tau),\nonumber\\
&\dot{x}_{2}=0,\nonumber\\
&\dot{y}_{2}=0.
\end{align}

\item [(ii)] \emph{No contact with moving capsule}: $x_2-x_1<\delta$, $y_2<0$ or $y_2>0$,
\begin{align}\label{movingcapsu_nocontactdelay}
&\dot{x}_{1}=y_{1},\nonumber\\
&\dot{y}_{1}=\alpha \cos(\omega \tau)+(x_{2}-x_{1})+2\zeta(y_{2}-y_{1})+u(\tau),\nonumber\\
&\dot{x}_{2}=y_{2},\nonumber\\
&\dot{y}_{2}=[-\sign(y_{2})-(x_{2}-x_{1})-2\zeta(y_{2}-y_{1})]/(\gamma).
\end{align}

\item[(iii)] \emph{Contact with stationary capsule}: $x_2-x_1>\delta$, $y_2=0$,
\begin{align}\label{nomovingcapsu_contactdelay}
&\dot{x}_{1}=y_{1},\nonumber\\
&\dot{y}_{1}=\alpha \cos(\omega
\tau)+(x_{2}-x_{1})+2\zeta(y_{2}-y_{1})-\beta(x_{1}-x_{2}-\delta)+
u(\tau),\nonumber\\
&\dot{x}_{2}=0,\nonumber\\
&\dot{y}_{2}=0.
\end{align}

\item[(iv)] \emph{Contact with moving capsule}: $x_2-x_1>\delta$, $y_2<0$ or $y_2>0$,
\begin{align}\label{movingcapsu_contactdelay}
&\dot{x}_{1}=y_{1},\nonumber\\
&\dot{y}_{1}=\alpha \cos(\omega
\tau)+(x_{2}-x_{1})+2\zeta(y_{2}-y_{1})-\beta(x_{1}-x_{2}-\delta)+
u(\tau),\nonumber\\
&\dot{x}_{2}=y_{2},\nonumber\\
&\dot{y}_{2}=[-\sign(y_{2})-(x_{2}-x_{1})-2\zeta(y_{2}-y_{1})+\beta(x_{1}-x_{2}-\delta)]/\gamma.
\end{align}

\end{itemize}
The resulting vector field governing the motion of $x_1$, $x_2$ will
be piecewise smooth and the phase space $\mathbb{R}^4$ for
{\color{black}$(x_1,y_1,x_2,y_2)$} will hence be divided into six different regions.
A detailed description of the properties of these phases and how to
develop the motion of the capsule system can be found in
\cite{liu2013modelling}. Then, based on the Filippov convention, the
equations of motion for the capsule system with the {\color{black}DFC} can be written in a compact form as follows
\cite{liu2013modelling}.
\begin{align}\label{capsule_delaymodel}
&\dot{x}_{1}=y_{1},\nonumber\\
&\dot{y}_{1}=+(x_{2}-x_{1})+2\zeta(y_{2}-y_{1})-H_{3}\beta(x_{2}-x_{1}-\delta)+\alpha
\cos(\omega \tau)
+u(\tau),\nonumber\\
&\dot{x}_{2}=y_{2}(H_{1}(1-H_{3})+H_{2}H_{3}),\nonumber\\
&\dot{y}_{2}=(H_{1}(1-H_{3})+H_{2}H_{3})[-\sign(y_{2})-(x_{2}-x_{1})-2\zeta(y_{2}-y_{1})+H_{3}\beta
(x_{1}-x_{2}-\delta)]/(\gamma),
\end{align}
where $H(\cdot)$ is the Heaviside function and functions
$H_{i}~(i=1,2,3)$ are defined as
\begin{align*}
&H_{1}=H( |(x_{2}-x_{1})+2\zeta(y_{2}-y_{1})|-1),\\
&H_{2}=H(|(x_{2}-x_{1})+2\zeta(y_{2}-y_{1})-\beta(x_{1}-x_{2}-\delta) |-1),\\
&H_{3}=H(x_{1}-x_{2}-\delta).
\end{align*}
In the present work, we will consider the capsule system
(\ref{capsule_delaymodel}) with the {\color{black}DFC}
$u(\tau)$ for $\tau\ge0$, where
\begin{equation}\label{delu4capsule}
u(\tau)=K\left(y_{1}(\tau-\tau_{d})-y_{1}(\tau)\right)\mbox{,\quad}\tau\geq0\mbox{,}
\end{equation}
is the {\color{black}DFC} {\color{black}with Pyragas-type delay feedback control form \cite{pyragas1992continuous}}. It should be pointed out that the
{\color{black}DFC} utilises real-time feedback relying on the
velocities of the inner mass at the present time $\tau $ and the
historical time $\tau-\tau_d$ only. In
Eq.~\eqref{delu4capsule}, the control gain $K\geq0$ determines the
coupling strength between the velocity output
$y_{1}(\tau-\tau_d)-y_{1}(\tau)$, and the time delay $\tau_{d}>0$ is
tunable (in contrast to a lag introduced by the control loop, which
is assumed to be zero here).
{\color{black}By fixing $\tau_d=T:=2\pi/\omega$, our purpose is to control the system to achieve the switching from the other attractors to a coexisting period-1 response.}
For zero gain, $K=0$,  $u(\tau)$ equals zero such that we have
purely harmonic excitation with
frequency $\omega$ and amplitude $\alpha$. 
When the {\color{black}DFC} is activated with some suitable
$K>0$, the trajectory of the original motion has a chance to be
stable on the desired periodic motion with the period, which equals
to the time delay $\tau_d$. After the original trajectory is
controlled to the desired attractor, the control signal $u$ becomes
very small, and the controlled system reverts to the
original system, {\color{black}which does not introduce or generate new attractors.}
{\color{black}Besides that, the DFC has many other advantages on controlling multistability. The DFC only needs the information of desired period of the targeting attractor, rather than the trajectory of the desired attractor. However, it should be noted that with introducing the DFC, the dimension of the controlled system becomes infinite, leading the dynamics of the controlled system \eqref{capsule_delaymodel} to be more sophisticated for analysis.}



\section{Numerical investigation of the capsule system under the {\color{black}DFC}}
In this section, we will {\color{black} numerically} study the effectiveness of the {\color{black}DFC} on controlling the multistability of the
vibro-impact capsule system. Meanwhile, some criterions, such as the
energy and convergence time, will be considered in our numerical
discussions to evaluate the control effect of the {\color{black}DFC} under different values of the control gain. Therefore, we
will consider two performance measures, which are the control energy
\begin{align}\label{energy}
E_{u}=\int_{\tau_{c}}^{\tau_{f}}u^{2}(\tau)d\tau,
\end{align}
where $\tau_c$ represents the time at which the control law is
activated, while $\tau_f$ ($\tau_f>\tau_c$) is a given final time of
the simulation, and the convergence time
\begin{align}\label{convergency}
\text{$T_\mathrm{conv}$:} \mathop{\arg\min}\limits_{\tau \in
[\tau_c, \tau_f]}
|| &[x_{1}(\tau),y_{1}(\tau),x_{2}(\tau),y_{2}(\tau)]^{T}\nonumber\\
&-[x_{d,1}(\tau),y_{d,1}(\tau),x_{d,2}(\tau),y_{d,2}(\tau)]^{T}||=\epsilon_{b},
\end{align}
where $\epsilon_{b}$ is the threshold distance between the
controlled trajectory and the desired attractor that normally should
be small,
$x_{d,i}$ and $y_{d,i}$, $(i=1,2)$, denote the displacements and
velocities of a desired trajectory. The convergence time
$T_\mathrm{conv}$ defined by Eq.~\eqref{convergency} measures the
time from switch-on of control to the time when the distance between
the controlled system and the desired trajectory reaches zero. The
first measure $E_{u}$ describes the expenditure of energy that is
consumed by the {\color{black}DFC} $u(\tau)$ over the time
interval $[\tau_c,\tau_f]$, where we favor lower values of $E_u$.
For the second measure, the convergence time $T_\mathrm{conv}$ given
by Eq.~\eqref{convergency}, describes the time it takes to reach the
desired attractor, where we also favor lower values of
$T_\mathrm{conv}$.
These two measures will help us evaluate the performance of the
{\color{black}DFC} \eqref{delu4capsule} for different
values of control gain $K$ based on the aspects of energy efficiency
and convergence speed and determine an optimal control gain $K$ for
controlling multistability.

\subsection{Control of multistability}

\begin{figure}[h!]
\centering
\includegraphics[width=\textwidth]{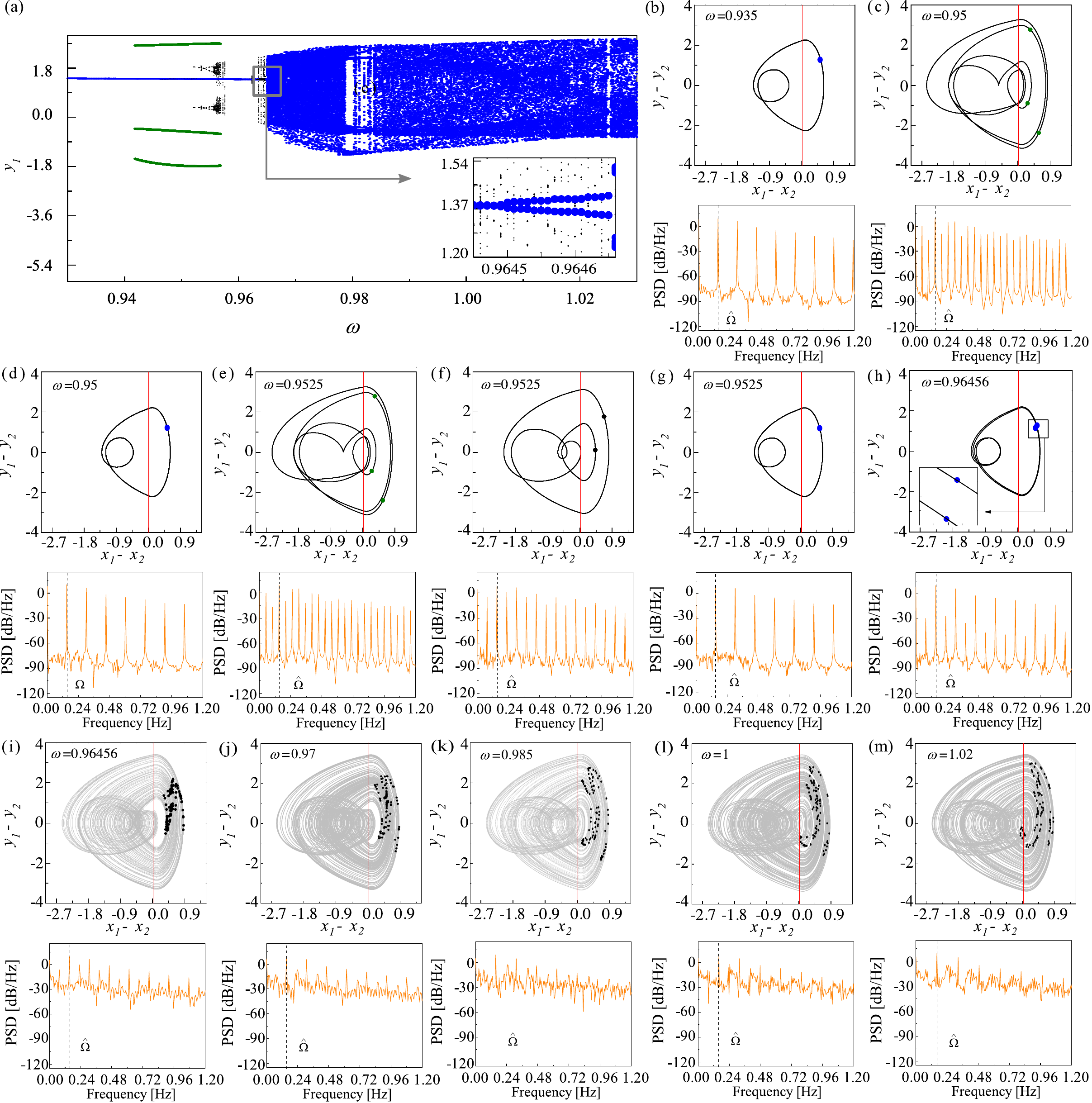}
\caption{(Colour online) (a) Bifurcation diagram of the vibro-impact
capsule system with respect to the excitation frequency
$\omega$, computed for $\alpha = 1.6$, $\zeta = 0.01$, $\delta=0.02$,
$\gamma=5$, $\beta = 15$. Blue, black and green dots denote different
coexisting attractors recorded in the computation. Additional
panels (b), (c)-(d), (e)-(g), (h)-(i), {\color{black}(j), (k), (l) and (m)} present the corresponding
attractors {\color{black}at} Poincar\'e sections {\color{black}and their power spectrum densities} for $\omega = 0.935$, $0.95$, $0.9525$, $0.96456$, {\color{black}$0.97$, $0.98$, $1$ and $1.02$, respectively}. Red vertical and {\color{black}black dash lines indicate} the location of the impact boundary and {\color{black}the driven frequency $\hat{\Omega}=\frac{\omega}{2\pi}$}, respectively.
}\label{bifur_nocontrol}
\end{figure}

\paragraph{{Bifurcation analysis of uncontrolled system for $\omega\in [0.93,0.98]$}}
Bifurcation analysis is carried out for the
vibro-impact capsule system \eqref{capsule_delaymodel} with $u=0$,
i.e., the vibro-impact system without controller. The results of a
parameter sweep are presented in Fig.~\ref{bifur_nocontrol} for
varying excitation frequency $\omega\in [0.93,0.98]$.
For excitation frequency $\omega$ around and above $1.02$, the relevant
detailed bifurcation analyses can be found in \cite{liu2015forward}. To construct the bifurcation diagram, zero initial condition, $(x_1,y_1,x_2,y_2)=(0,0,0,0)$, was firstly used. Then both of backward and forward sweeping of bifurcation parameter $\omega$ was locally performed for coexisting attractors.
In the simulation, $350$ periods of external excitation
were used to obtain each value of the bifurcation parameter
$\omega$. The outputs for the first 300 cycles were discarded as transients,
whereas the values of velocity of inner mass $y_{1}$ at
$\tau=\frac{2n\pi}{\omega}$, $n=301,\ldots,350$ are included in a
  scatter plot in Fig.~\ref{bifur_nocontrol}, generating a bifurcation diagram in the
  $(\omega,y_1)$-plane. We observe several regions with visibly different qualitative behaviour of
  system \eqref{capsule_delaymodel} with $u=0$, {\color{black}and the regularity of typical capsule dynamics can be confirmed by the power spectrum density of capsule's velocity.}
\begin{itemize}
\item For $\omega\in [0.93, 0.94185]$ the system shows a period-1 motion as the only attractor reached. {\color{black} The top panel of}
    Fig.~\ref{bifur_nocontrol}(b) shows the period-$1$ trajectory, {\color{black}which is a regular motion with sharp peaks at the driven frequency $\hat{\Omega}=\frac{\omega}{2\pi}$ and its super-harmonics, as shown in the lower panel of Fig.~\ref {bifur_nocontrol} (b).}
\item For $\omega\in(0.94185,0.95215]$ two coexisting
  attractors are found, of which one is the period-1 response
  (blue) continuing the attractor from smaller $\omega$, and
  the other is a period-3 (green) response. {\color{black}From the power spectrum density presented in Fig.~\ref {bifur_nocontrol}(c) and (d), the period-1 and period-3 responses also have sharp peaks at the driven frequency $\hat{\Omega}$, and their sub- and super-harmonics, with ignorable power.}
\item  For $\omega\in(0.95215,0.9561]$
  three coexisting motions are {\color{black}observed}.
  For example, at $\omega=0.9525$, the system has a period-1 (blue), a
  period-3 (green) and a period-2 (black) responses (see {\color{black}Fig.~\ref{bifur_nocontrol}(e)-(g)}). The period-$2$ attractor (black in Fig.~\ref{bifur_nocontrol})
  undergoes a sequence of bifurcations near the large-$\omega$ end of this region.
\item For $\omega>0.9561$ the period-3 attractor {\color{black}does not exist anymore}. At $\omega\approx0.96453$,
the period-$1$ response undergoes a period doubling, and then
becomes to a period-$2$ motion.
  The trajectory of the emerging period-2 response at $\omega=0.96456$ is shown in {\color{black}Fig.~\ref{bifur_nocontrol}(h)}, and the coexisting chaotic motion is {\color{black}displayed in Fig.~\ref{bifur_nocontrol}(i)}.
  The range of this period-2 motion is small, giving way to
  the chaotic motion for $\omega>0.96469$. {\color{black}The detailed dynamics of this chaotic motion can be seen from Fig.~\ref{bifur_nocontrol}(i)-(m), where their power spectrum densities are distributed in a wide frequency region.}
\end{itemize}

\begin{figure}[h!]
\centering
\includegraphics[width=\textwidth]{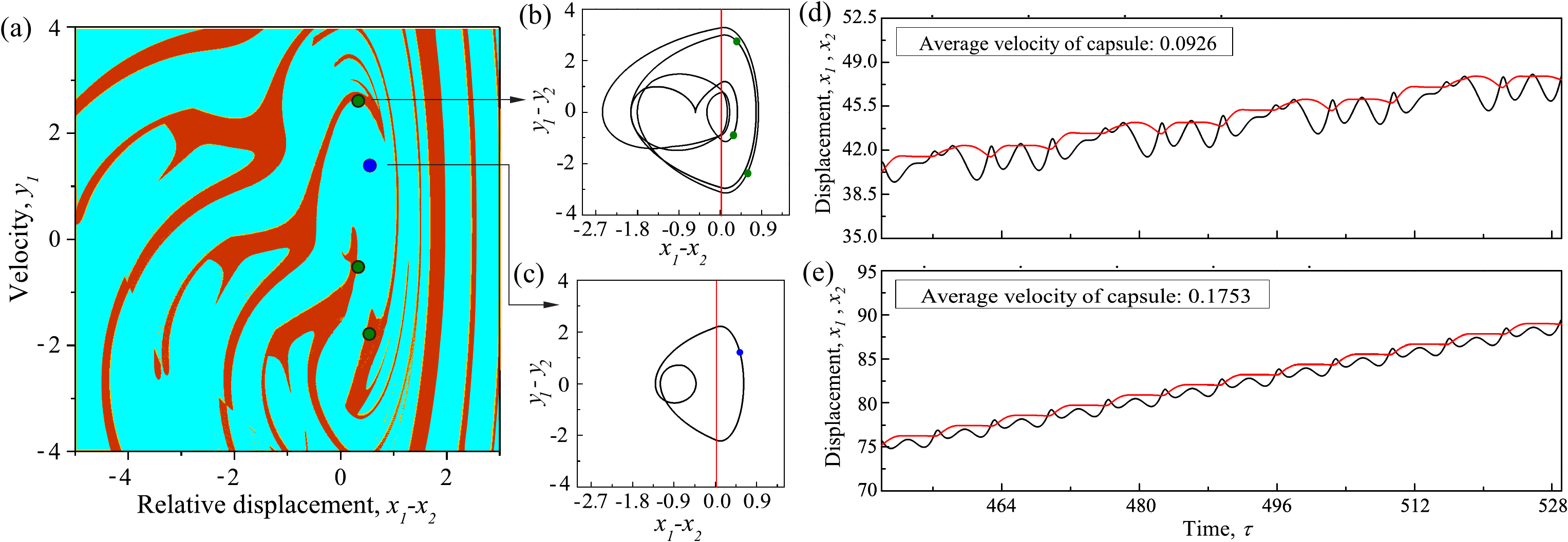}
\caption{(Colour online) (a) Basins of attraction for the two
attractors of the vibro-impact capsule system calculated for $\alpha = 1.6$, $\zeta
= 0.01$, $\delta=0.02$, $\gamma=5$, $\beta = 15$ and $\omega=0.95$;
  (b) period-3 trajectory (green dots with red basin); (c) period-1 trajectory (blue dot with cyan basin), where vertical red lines indicate the location of the impact boundary. (d) Time histories of displacements of the mass, $x_1$ (black line) and the capsule, $x_2$ (red line) for the period-3 response. (e) Time histories of displacements of the mass, $x_1$ (black line) and the capsule, $x_2$ (red line) for the period-1 response.
}\label{bistability}
\end{figure}

\paragraph{{Basins of attraction and attractor properties for bistable uncontrolled system at $\omega=0.95$}} As shown in the bifurcation analysis in Fig.~\ref{bifur_nocontrol},
the system at $\omega=0.95$ has two coexisting attractors. The
basins of attraction of two coexisting periodic responses are
presented in Fig.~\ref{bistability} (a), of which one is the
period-$3$ attractor shown in Fig.~\ref{bistability} (b), and the
other one is the period-$1$ attractor shown in
Fig.~\ref{bistability} (c). Figure~\ref{bistability}(d,e) show that
the average asymptotic progression velocities
$\lim_{\tau\to\infty}(\omega/(2n_{p}\pi))\int_\tau^{\tau+2n_{p}\pi/\omega}
y_2(s)\mathrm{d}s$, $n_p=1,3$, of the capsule of these two
coexisting motions are different. Specifically, the average velocity
of capsule of the period-3 response is $0.0926$, but the average
velocity of capsule of the period-$1$ response is $0.1753$, which is
almost twice that of the period-$3$ response.
Undoubtedly, the capsule presenting the period-1 response progresses
further in the same time period and has a faster speed than a
capsule showing the period-$3$ response. Hence, in this scenario, it
may be desirable to implement a control $u$ that steers system
\eqref{capsule_delaymodel} from its period-$3$ response with lower
average progression speed $ y_2$ to its period-$1$
response with higher speed. 

\begin{figure}[h!]
\centering
\includegraphics[width=\textwidth]{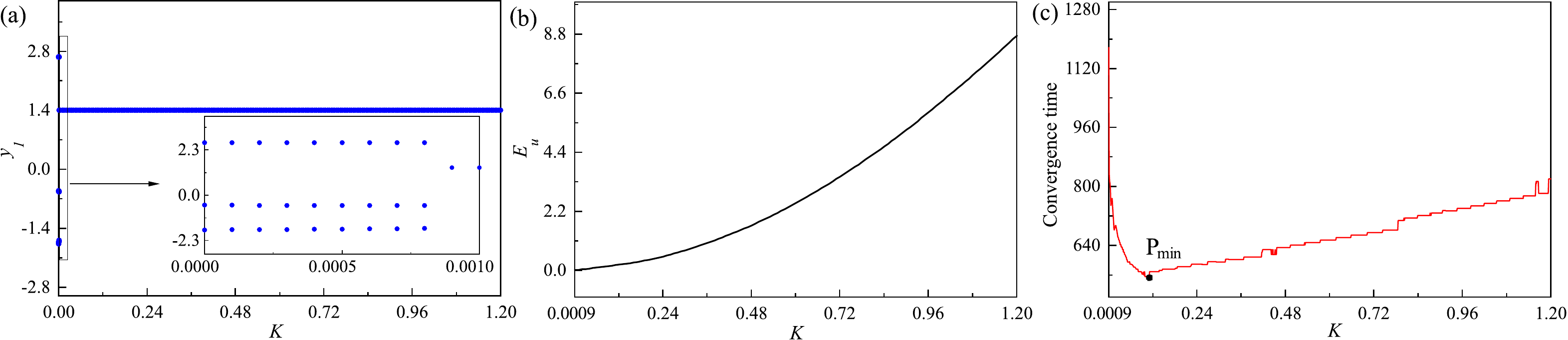}
\caption{(Colour online) {\color{black}(a) Bifurcation diagram, (b) control energy $E_u$ and (c) convergence time of the vibro-impact capsule system with the {\color{black}DFC}, where the control gain $K$ is considered as a branching parameter calculated for $\alpha = 1.6$, $\zeta = 0.01$, $\delta=0.02$, $\gamma=5$, $\beta = 15$, $\omega=0.95$ and $\tau_d=\frac{2\pi}{\omega}$.}
}\label{delaybifur_k}
\end{figure}

\paragraph{Effect of the {\color{black}DFC} on the attractors at $\omega=0.95$} The following
analysis studies the {\color{black}DFC} \eqref{delu4capsule} for 
controlling multistability of 
the vibro-impact capsule system \eqref{capsule_delaymodel}. In order to demonstrate the feasibility of the {\color{black}DFC} on achieving the above target, we present the bifurcation
analysis of the controlled system \eqref{capsule_delaymodel} for
varying control gain $K$ in Fig.~\ref{delaybifur_k} (a). In
Fig.~\ref{delaybifur_k} (a), there are 300 cycles of
external excitation used to obtain the bifurcation diagram, and
only the last 30 values of $y_1$ were plotted in the bifurcation
diagram for each value of the parameter $K$ to ensure the steady
state response. Since there are two coexisting periodic
responses at $\omega=0.95$ and among them the period-1 is desired,
the {\color{black}DFC} \eqref{delu4capsule} with
$\tau_d=\frac{2\pi}{\omega}$ was applied to achieve the switching
from the period-$3$ response to the period-1 response.
Fig.~\ref{delaybifur_k}(a) shows that {\color{black}DFC} has a
large range for the control gain $K$ for which it reaches the
control target (the period-$1$ response) from the period-$3$
response. When $K\ge 0.0009$, the period-3 response can always be
controlled to present period-1 motion. Since the feasible range of
the control gain $K$ is large, we have to consider some other
performance criteria, such as the energy expenditure $E_u$, given in
\eqref{energy}, and convergence time $T_\mathrm{conv}$, given in
\eqref{convergency}, to determine a suitable value of the control
gain $K$. Fig.~\ref{delaybifur_k} (b) and (c) show $E_u$ and
$T_\mathrm{conv}$ for varying control gain $K$. Initially the energy
expenditure $E_{u}$ stays at a lower level which is smaller than
$0.5109$, and increase slowly, when $K\in [0.0009,0.24]$. Then,
after $0.24$, $E_u$ starts to increase exponentially, which should
not be considered as the suitable values for controlling the above
multistability scenario. The convergence time $T_\mathrm{conv}$,
shown in Fig.~\ref{delaybifur_k}(c) for varying $K$, has a minimum
near $K=0.11$ ($\min T_\mathrm{conv}\approx551.459$).
Since $K=0.11$ is in the range  $[0.0009,0.24]$, where $E_u$ dose not witness significant change, 
one may consider $K=0.11$ as the optimal value for control gain $K$
(when weighting the measures $E_u$ and $T_\mathrm{conv}$
approximately equally).

\begin{figure}[h!]
\centering
\includegraphics[width=\textwidth]{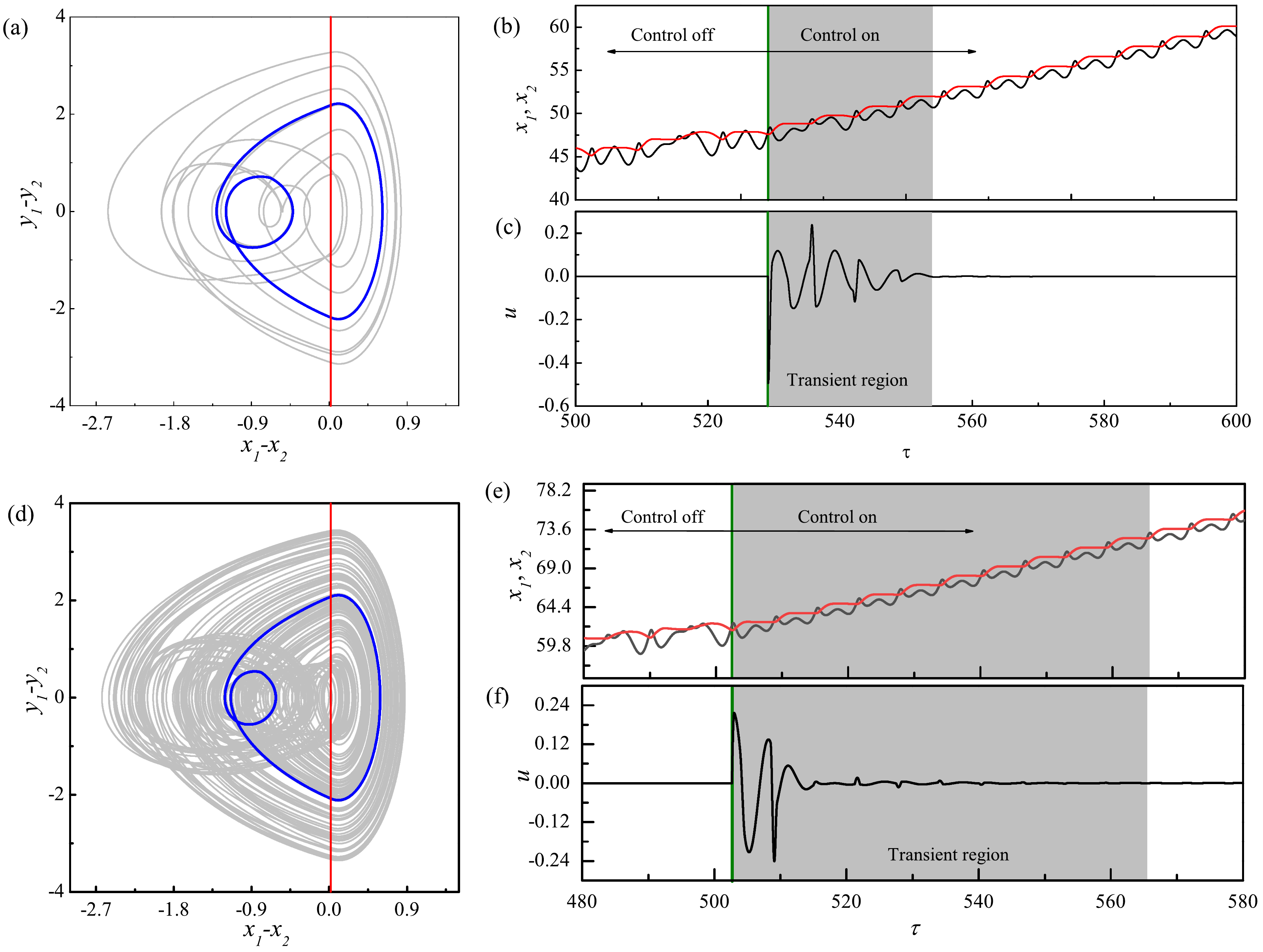}
\caption{(Colour online) {\color{black} (a) Phase trajectory, (b) displacement and (c) control signal of the period-3 motion before and after the DFC is switched on, computed for $\alpha = 1.6$, $\zeta = 0.01$, $\delta=0.02$, $\gamma=5$, $\beta = 15$, $\tau_d=\frac{2\pi}{\omega}$, $K=0.11$, and $\omega=0.95$. (d) Phase trajectory, (e) displacement and (f) control signal of the chaotic motion before and after the DFC is switched on, computed for $\alpha = 1.6$, $\zeta = 0.01$, $\delta=0.02$, $\gamma=5$, $\beta = 15$, $\tau_d=\frac{2\pi}{\omega}$, $K=0.11$, and $\omega=1$. Grey and blue trajectories are without and with the control, red and green vertical lines denote the impact boundary and the time that control is applied, respectively, and grey box indicates the transient region of the controlled system.}
}\label{bifur_controldetails}
\end{figure}

Following the above analysis, $K=0.11$ is chosen as
the ideal control gain of the delay feedback control
for stabilising the vibro-impact capsule system to the period-1
response. Fig.~\ref{bifur_controldetails} shows the
varying of the mass and capsule position ($x_1$ and $x_2$) and the
control signal $u$ with respect to the time history. The control
gain $K$ was set from $0$ to $0.11$ after the $81$'th period of the
excitation, with the effect that the period-$3$ response was
perturbed, steering the system towards the desired period-$1$
response.
{\color{black}Then the system converges to period-1 response at $\tau=551.459$.}
 The details of the period-$1$ response can be discovered
from {\color{black}the phase plane}, as shown
in Fig.~\ref{bifur_controldetails} {\color{black}(a)}. In addition, the
control signal $u$ converges to zero asymptotically, when the
control stabilises the vibro-impact capsule system on the period-$1$
response.
{\color{black}Similarly, when $\omega=1$, the chaotic response can be controlled to the unstable period-1 response by the DFC with the same control gain as shown in Fig.~\ref{bifur_controldetails} (d-f).}

\subsection{Variation of the delay}
In order to investigate the {\color{black}DFC's performance under the variation of delay time $\tau_d$,} we will introduce two
measures below. The first one is the
\emph{Maximum Velocity Difference} given as
\begin{align}\label{distan_delay}
\operatorname{mvd}(T,\tau_d):= \max_{\tau>\tau_\mathrm{stab}}
|y_{1}(\tau-T)-y_{1}(\tau)|,
\end{align}
{\color{black}where $\tau_{stab}$ denotes the time when transient response terminates}, and $T=2\pi/\omega$
is one period of the driving excitation. The second measure is a
quantitative characterisation of the invasiveness of the {\color{black}DFC} \cite{mfoumou2019computational} written as
\begin{align}\label{quality_delay}
q_{c}(\tau_d,K):=  \lim_{\delta \tau  \to +\infty} \frac{1}{\delta
\tau} \int_{\tau_c}^{\tau_c+\delta \tau} K^2 \langle
y_{1}(\tau-\tau_d)-y_{1}(\tau)  \;  ,  \;
y_{1}(\tau-\tau_d)-y_{1}(\tau) \rangle d \tau.
\end{align}
In our simulation, we chose $\tau_\mathrm{stab}$ as the $100$th
period of the external excitation after the control gain $K$ was
increased from $0$. In our observation the transients of the
controlled system have already decayed after this time. When the measure
$\operatorname{mvd}(T,\tau_d)$ is zero, the
feedback control with delay time $\tau_d$ asymptotically (after
$\tau_\mathrm{stab}$) achieves the period-$1$ response. The measure $q_c$ in \eqref{quality_delay} presents the average power used to control the capsule system over a certain delay time $\tau_d$.

\begin{figure}[h!]
\centering
\includegraphics[width=\textwidth]{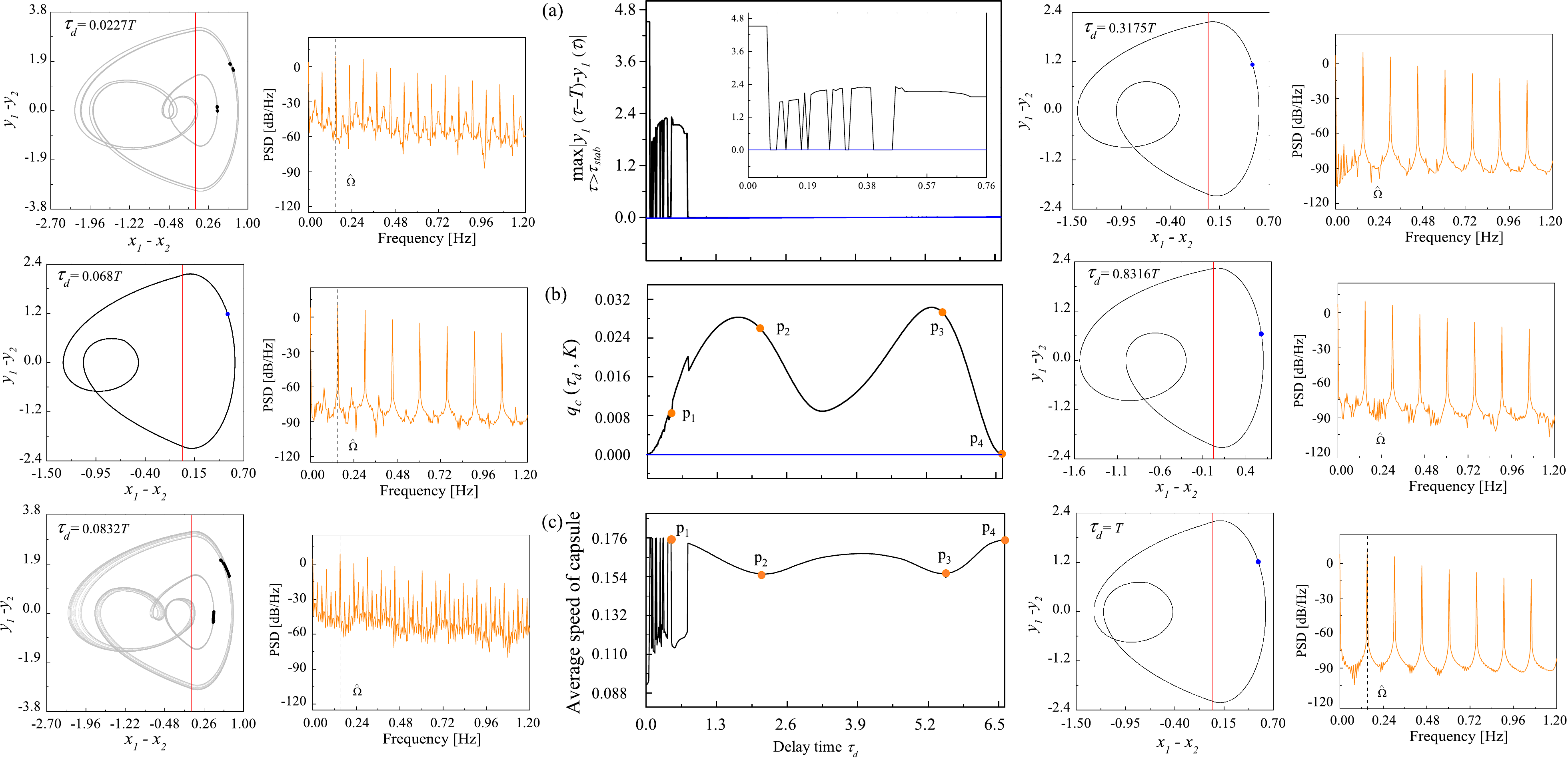}
\caption{(Colour online) {\color{black}(a) Variation of the maximum absolute difference between $y_{1}(\tau-T)$ and $y_{1}(\tau)$, (b) the measure $q_c$ in Eq.~\eqref{quality_delay} and (c) the average velocity of the capsule system as functions of the delayed time $\tau_d$, computed for $\alpha = 1.6$, $\zeta = 0.01$, $\delta=0.02$, $\gamma=5$, $\beta = 15$, $\omega=0.95$, $K=0.11$ and $T=\frac{2\pi}{\omega}$, where $\rm{p}_1$, $\rm{p}_2$, $\rm{p}_3$ and $\rm{p}_4$ represent typical period-1 responses for $\tau_d=0.068T$, $\tau_d=0.3175T$, $\tau_d=0.8316T$ and $\tau_d=T$, respectively. Additional panels demonstrate phase trajectories and power spectrum densities of the system recorded for $\tau_d=0.0227T$ (chaotic), $\tau_d=0.068T$ (period-1), $\tau_d=0.0832T$ (chaotic), $\tau_d=0.3175T$ (period-1), $\tau_d=0.8316T$ (period-1) and $\tau_d=T$ (period-1). Blue horizontal lines indicate zero value, red vertical lines denote the location of the impact boundary, and vertical dash lines indicate the driven frequency $\hat{\Omega}=\frac{\omega}{2\pi}$.}
}\label{bifur_differdelay}
\end{figure}

Our main {\color{black}purpose} in this section {\color{black}is to study} how the delay time
$\tau_d$ in the DFC \eqref{delu4capsule}
affects the measures $\operatorname{mvd}(T,\tau_d)$ and $q_c$ for
the controlled system \eqref{capsule_delaymodel}.
Fig.~\ref{bifur_differdelay} shows the dynamic response of system
\eqref{capsule_delaymodel} for a parameter sweep of the delay time
$\tau_d$. The measure $\operatorname{mvd}(T,\tau_d)$ in
\eqref{distan_delay} shows that there are many delay times
$\tau_d\in[0,T]$ for which system \eqref{capsule_delaymodel} reaches
a period-$1$ response from the period-3 response, such as
$\tau_d=0.068T$, $\tau_d=0.3175T$, $\tau_d=0.8316T$ and $\tau_d=T$
as shown in Fig.~\ref{bifur_differdelay} {\color{black}(c) and (g-i), where all the periodic motions have sharp peaks at the driven frequency and their sub- and super-harmonics}. Some
chaotic responses were also observed at some values of the delay time
$\tau_d$, such as $\tau_d=0.0227T$ and $\tau_d= 0.0832T$ as shown in
Fig.~\ref{bifur_differdelay} (a) and {\color{black}(e), at where distribute their energies in a wide frequency region and present irregular dynamics}.

The measure $q_c$ in Eq.~\eqref{quality_delay} quantifies the control
performance of the DFC with respect to the delay
time $\tau_d$. {\color{black}As a detailed study was carried out by P\'aez Ch\'avez et al. \cite{chavez2016path}, we will not study progression optimisation for the capsule system in the present work.}
As shown in Fig.~\ref{bifur_differdelay}{\color{black}(a) and (b)}, the relevant
$q_{c}(\tau_d,K)$ at $\rm{p}_1~(\tau_d=0.068T)$,
$\rm{p}_2~(\tau_d=0.3175T)$, $\rm{p}_3~(\tau_d=0.8316T)$ and
$\rm{p}_4~(\tau_d=T)$  are $0.0085$, $0.026$, $0.0293$ and
$9.438\times10^{-5}$, respectively.
It should be pointed out that, among these delay times,
$q_{c}(\tau_d,K)$ at $\tau_d=0.068T$ is the lowest except for
$\tau_d=T$, and the relevant DFC at this delay
time can stabilise the system at the period-$1$ response.
{\color{black}On the other hand, the average speed of the capsule at $\rm{p}_1$ is also the fastest around $0.175$ as shown in Fig.~\ref{bifur_differdelay}(c), and the rest, $\rm{p}_2$, $\rm{p}_3$ and
$\rm{p}_4$ all present high efficient progression speeds.}
In fact, it is well-known that the dimension of the time-delayed system
will be increased with the increasing delay time in the delayed
arguments on the right hand side, such that the analysis of
the dynamical properties becomes more complex. {\color{black}However, the
above analysis shows that DFC
\eqref{delu4capsule} with a small delay time $\tau_d$ is able to
stabilise the capsule system on a period-$1$ response using a small cost of input while keeping high efficient progression speed. Then system \eqref{capsule_delaymodel} has a low dimension for small
$\tau_d$, allowing us to carry out a more detailed bifurcation analysis by using
the numerical continuation techniques in the next section.}

\section{Numerical continuation study of the capsule system with the DFC}
\label{sec-delaycontroller}

This section will investigate the periodic response of the capsule
model using numerical continuation methods, taking into account the
time-delayed feedback controller described in previous sections. As
pointed out before, the main challenge in this study is the fact
that, to the best of our knowledge, there is at the moment no
specialized software package for numerical continuation of
piecewise-smooth delay differential equations. For this reason, the
path-following analysis to be presented in this section will be
implemented using the approximate numerical approach proposed in
\cite{joseph2020}, which will be briefly introduced below.

\subsection{Approximate equations for numerical continuation of piecewise-smooth
DDEs}\label{sec-app-dde}

The numerical framework for the path-following analysis of the
capsule system under the {\color{black}DFC} is based on the
\emph{chain method} \cite{repin65}, improved by a higher-order
approximation scheme using a finite sequence of Taylor expansions as
follows. Consider a system of delay differential equations (DDEs)
with constant delay $\tau_{d}>0$
\begin{equation}\label{eq-DDE}
\dot{q}(\tau)=f(\tau,q(\tau),q(\tau-\tau_{d})),
\end{equation}
where
$f:\mathbb{R}\times\mathbb{R}^d\times\mathbb{R}^d\rightarrow\mathbb{R}^d$
is sufficiently smooth. Choose an $N\in\nat$ suitably large and
introduce the grid points $\tau_{i}:=i\frac{\tau_{d}}{N}$,
$i=0,\ldots,N$. Furthermore, define the finite sequence of functions
$v_{i}(\tau):=q(\tau-\tau_{i})$ for all $\tau\geq0$, $i=0,\ldots,N$.
In this setting, via an $M$th-order Taylor expansion we obtain
\begin{equation}\label{eq-Taylor-disc}
v_{i-1}(\tau)=q\left(\tau-\left(\tau_{i}-\frac{\tau_{d}}{N}\right)\right)=
v_{i}\left(\tau+\frac{\tau_{d}}{N}\right)=
\sum^{M}_{k=0}\frac{1}{k!}v^{(k)}_{i}(\tau)\left(\frac{\tau_{d}}{N}\right)^k+
\bigo\left(\left(\frac{\tau_{d}}{N}\right)^{M+1}\right),\mbox{
}\mbox{ }\mbox{ and }
\end{equation}
\begin{equation}\label{eq-DDE-insert}
\dot{v}_{0}(\tau)=f(\tau,v_{0}(\tau),v_{N}(\tau)),
\end{equation}
with $\tau\geq0$, $i=1,\ldots,N$ and $M\geq1$. Upon neglecting the
$\bigo$-terms, a system of $dN$ differential equations of order $M$
is derived from \eqref{eq-Taylor-disc}. This allows us to
approximate a piecewise-smooth dynamical system of dimension $d$
with constant delay via a piecewise-smooth system of first-oder ODEs
of dimension $d(NM+1)$ for large $N$. In this way, the resulting
model can be studied in the context of hybrid dynamical systems
\cite{thota2008tc}.

\subsection{Mathematical formulation of the time-delayed capsule system for path-following
analysis}

Let us choose $M=2$ in the approximation scheme introduced before
and denote by $z=(x_{r},v_{r},r,s,v_{0},$
$\ldots,v_{N},w_{1},\ldots,w_{N})^T\in\mathbb{R}^{2N+5}$ and
$\lambda=(\omega,\alpha,\zeta,\delta,\gamma,\beta,K,\tau_{d})\in\left(\mathbb{R}^+_{0}\right)^8$
the state variables and parameters of the system, respectively. We
consider the coordinate transformation $x_{r}=x_{1}-x_{2}$,
$v_{r}=y_{1}-y_{2}$, $v_{0}=y_{1}$. In this setting, an
approximation of the capsule model with {\color{black}the DFC} via
a system of piecewise-smooth ODEs is given by
\begin{align}
\dot{z}&=\left(\setstretch{1.25}\begin{array}{c}
\tau_{d}v_{r}\\
\tau_{d}\left(\alpha s+\overline{u}-x_{r}-2\zeta
v_{r}-H_{\mathrm{k_{2}}}\beta(x_{r}-\delta)-\abs{H_{\mathrm{vel}}}\left(x_{r}+2\zeta
v_{r}+H_{\mathrm{k_{2}}}\beta(x_{r}-\delta)-H_{\mathrm{vel}}\right)/\gamma\right)\\
r+\omega\tau_{d}s-r\left(r^2+s^2\right)\\
s-\omega\tau_{d}r-s\left(r^2+s^2\right)\\
\tau_{d}\left(\alpha s+\overline{u}-x_{r}-2\zeta
v_{r}-H_{\mathrm{k_{2}}}\beta(x_{r}-\delta)\right)\\
\left(w_{i}\right)_{i=1,\ldots,N}\\
\left(2N^2\left(v_{i-1}-v_{i}-\dfrac{1}{N}w_{i}\right)\right)_{i=1,\ldots,N}
\end{array}\right)\nonumber\\
&=f_{\text{\tiny
CAP}}(z,\lambda,H_{\mathrm{k_{2}}},H_{\mathrm{vel}}),
\label{eq-hybsys-dde-app}
\end{align}
where $w_{i}$ are auxiliary variables representing the first
derivative of $v_{i}$, $1\leq i\leq N$, and
$\overline{u}(\tau)=K(v_{N}(\tau)-v_{0}(\tau))$, $\tau\geq0$, stands
for the approximate delay feedback control signal. Notice that a
time re-scaling $\tau\leftarrow\tau/\tau_{d}$ has been adopted in
the approximating system \eqref{eq-hybsys-dde-app}. In this way, the
approximation of the history $q(\tau-t)$ for $t\in[0,\tau_d]$ given
by the Taylor expansion \eqref{eq-Taylor-disc} is restricted to the
unit interval, for any delay $\tau_{d}$. Moreover, for the
approximating system \eqref{eq-hybsys-dde-app} one has that
$v_{0}(\tau)=y_{1}(\tau)$ and
\begin{equation}\label{eq-uis}
v_{i}(\tau)\approx y_{1}(\tau-\tau_{i}),\mbox{ }\mbox{ for all
}\mbox{ }\tau\geq0,\mbox{ }\mbox{ }\mbox{
}\tau_{i}=\frac{i}{N},\mbox{ }\mbox{ }\mbox{ }i=1,\ldots,N.
\end{equation}

The approximation scheme introduced here can be considered as a
special case of the pseudo-spectral approximation of DDEs by ODEs
proposed in \cite{breda2016pseudospectral}. Nevertheless, this
spectral approximation presents no advantage over the low-order
approximation \eqref{eq-uis}, since the history segment is only
differentiable once with Lipschitz continuous derivative whenever a
discontinuity boundary (i.e.\ impact with the secondary spring
$k_{2}$, transition from forward to backward capsule motion and so
on) is crossed, and therefore the second-order approximation
introduced in the present work has the most suitable order $M$.

Moreover, note that the state variables in the formulation above do
not include a coordinate for the capsule position. This information,
however, can be recovered from model \eqref{eq-hybsys-dde-app} as
follows:
\begin{equation*}
x_{c}(\tau):=x^*_{c}+\int\limits_{0}^{\tau}(v_{0}(s)-v_{r}(s))\,ds,
\end{equation*}
where $x^*_{c}\in\re$ represents the position of the capsule
(coordinate $x_{2}$ in the original model) at $\tau=0$.

On the other hand, the symbols $H_{\mathrm{k_{2}}}$ and
$H_{\mathrm{vel}}$ appearing in the approximate system
\eqref{eq-hybsys-dde-app} are discrete variables defining the
operation modes of the system, according to the rules:
\begin{equation}\label{eq-Hk2}
H_{\mathrm{k_{2}}}(x_{r})=\begin{cases}
1, & x_{r}-\delta\geq0,\mbox{ }\mbox{ }\mbox{(contact with secondary spring $k_{2}$)},\\
0, & x_{r}-\delta<0,\mbox{ }\mbox{ }\mbox{(no contact)},
\end{cases}
\end{equation}

\begin{equation}\label{eq-Hvel}
H_{\mathrm{vel}}(v_c,f_\mathrm{mc})=\begin{cases} \phantom{-}0, &
v_{c}=0\mbox{ }\mbox{ and }\mbox{ }\abs{f_\mathrm{mc}}\leq1
,\mbox{ }\mbox{ }\mbox{(capsule stationary)},\\
\phantom{-}1, & v_{c}>0\mbox{ }\mbox{ or}\mbox{ }\left(v_{c}=0\mbox{
}\mbox{ and }\mbox{ }f_\mathrm{mc}>1\right),\mbox{
}\mbox{ }\mbox{(forward motion)},\\
-1, & v_{c}<0\mbox{ }\mbox{ or}\mbox{ }\left(v_{c}=0\mbox{ }\mbox{
and }\mbox{ }f_\mathrm{mc}<-1\right),\mbox{ }\mbox{ }\mbox{(backward
motion)},
\end{cases}
\end{equation}
where
\begin{align*}
v_{c}&=y_{1}-v_{r}&&\mbox{(capsule velocity $y_{2}$ in the original
  model), and}\\
f_\mathrm{mc}&=x_{r}+2\zeta
v_{r}+H_{\mathrm{k_{2}}}\beta(x_{r}-\delta)&&\mbox{(force acting on
the capsule from the internal mass).}
\end{align*}
Therefore, the capsule changes between stationary and forward
(backward) motion, whenever the force $f_{\mathrm{mc}}$ becomes
greater than $1$ (or smaller than $-1$). 
For the numerical implementation, the discrete variables defined in
\eqref{eq-Hk2}--\eqref{eq-Hvel} will be used to identify the
specific operation mode of the capsule.  Every operation mode will
be associated to a pair $\left\{\Sigma,\Delta\right\}$, where
$\Sigma\in\left\{\mbox{NC},\mbox{Ck2}\right\}$ (no contact or
contact with secondary spring $k_{2}$, respecitvely) and
$\Delta\in\left\{\mbox{Vc0},\mbox{Vcp},\mbox{Vcn}\right\}$ (capsule
stationary, forward motion, backward motion). For instance, the
operation mode $\left\{\mbox{Ck2},\mbox{Vcp}\right\}$ means that the
capsule is moving forward with the internal mass in contact with the
spring $k_{2}$ and so on. In this way, the capsule system can
operate under 6 different modes, as listed in Table \ref{tab-modes}.

\begin{table}[h]
\caption{Operation modes of the capsule system and the corresponding
values of the discrete variables $H_{\mathrm{k_{2}}}$ and
$H_{\mathrm{vel}}$ defined in \eqref{eq-Hk2}--\eqref{eq-Hvel}.}
\begin{center}
\setstretch{1.2}\begin{tabular}{|c|c|c|c|}\hline \textbf{Operation
mode} & \textbf{Description} & $H_{\mathrm{k_{2}}}$ &
$H_{\mathrm{vel}}$\\
\hline
$\left\{\mbox{NC},\mbox{Vc0}\right\}$ & No contact with $k_{2}$, capsule stationary & 0 & 0\\
\hline
$\left\{\mbox{NC},\mbox{Vcp}\right\}$ & No contact with $k_{2}$, forward motion & 0 & 1\\
\hline
$\left\{\mbox{NC},\mbox{Vcn}\right\}$ & No contact with $k_{2}$, backward motion & 0 & -1\\
\hline
$\left\{\mbox{Ck2},\mbox{Vc0}\right\}$ & Contact with $k_{2}$, capsule stationary & 1 & 0\\
\hline
$\left\{\mbox{Ck2},\mbox{Vcp}\right\}$ & Contact with $k_{2}$, forward motion & 1 & 1\\
\hline
$\left\{\mbox{Ck2},\mbox{Vcn}\right\}$ & Contact with $k_{2}$, backward motion & 1 & -1\\
\hline
\end{tabular}
\end{center}
\label{tab-modes}
\end{table}

\subsection{Continuation of periodic orbits}

In this section, a numerical continuation study of the delayed
capsule model will be carried out using the approximating system
\eqref{eq-hybsys-dde-app} (with $N=30$), implemented in the
continuation platform COCO \cite{dankowicz2013recipes}. For this
purpose, the COCO-toolbox `hspo' will be extensively used, which
implements a segment-specific discretization strategy in the
framework of multisegment boundary-value problems, suitable for
numerical continuation and bifurcation detection of periodic
solutions of piecewise-smooth dynamical systems. On the other hand,
all numerical simulations obtained via direct numerical integration
will be generated from the original DDE model
\eqref{capsule_delaymodel} using the MATLAB solver `\texttt{dde23}',
in combination with its built-in event location routines
\cite{shampine2000,shampine2001}. In this way, we will be able to
detect accurately collisions with the system's discontinuity
boundaries (e.g.\ impact with the secondary spring $k_{2}$,
transition from forward to backward capsule motion and so on).

\begin{figure}[H]
\centering \psfrag{T}{\normalsize Contact
time}\psfrag{om}{\normalsize$\omega$}
\psfrag{xr}{\normalsize$x_{r}$}\psfrag{vr}{\normalsize$v_{r}$}\psfrag{t}{\normalsize$t$}
\psfrag{xc}{\normalsize$x_{2}$}
\psfrag{a}{$\omega=0.95$}\psfrag{b}{$\omega=0.96471$}\psfrag{c}{$\omega=0.97$}
\includegraphics[width=0.8\textwidth]{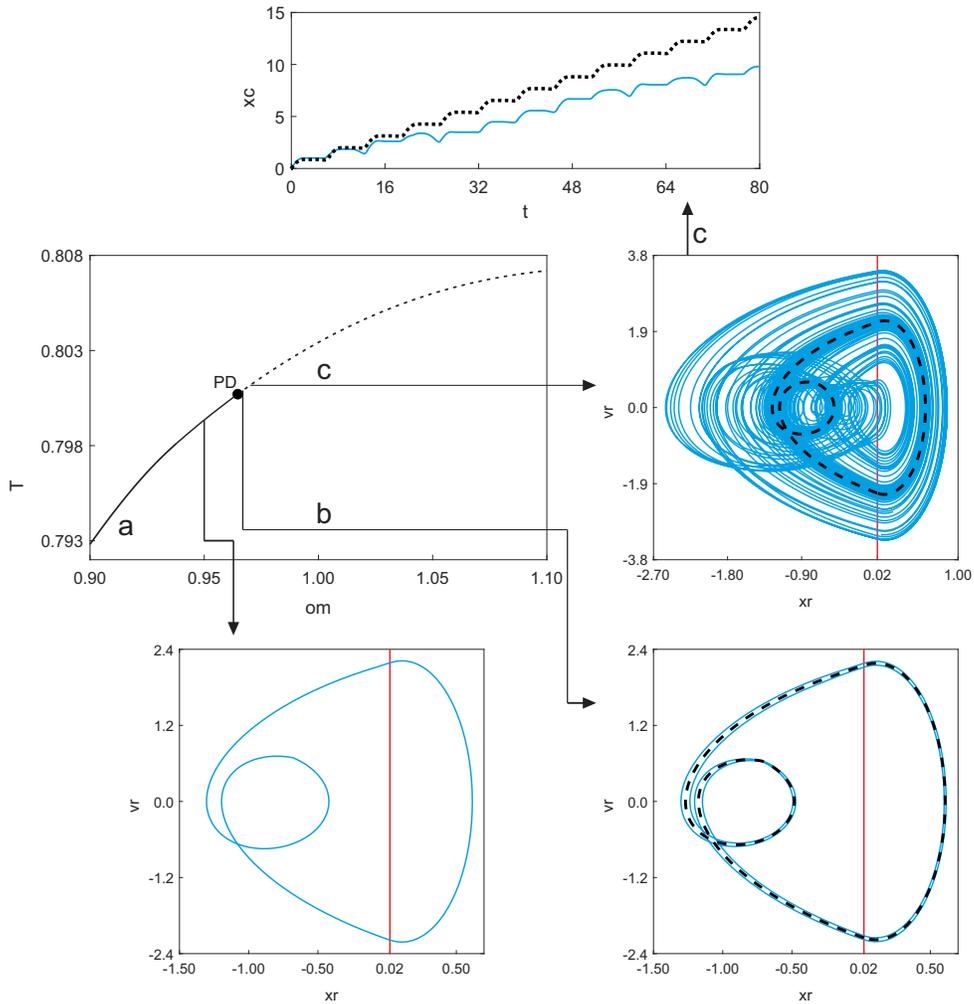}
\caption{Numerical continuation of period-1 orbits of the capsule
system \eqref{capsule_delaymodel} with respect to the excitation
frequency $\omega$, with $u=0$ (i.e.\ without delay control), and
for the parameter values $\alpha=1.6$, $\zeta=0.01$, $\delta=0.02$,
$\gamma=5$, $\beta=15$. The vertical axis presents the time the
oscillating mass is in contact with the secondary spring $k_{2}$.
The label PD stands for a period-doubling bifurcation of limit
cycles detected at $\omega\approx0.96453$. Branches of stable and
unstable solutions are marked by solid and dashed lines,
respectively. Lateral and lower panels depict phase plots
corresponding to the test points shown in the bifurcation diagram.
The location of the impact boundary $x_{r}=\delta$ is represented by
a vertical red line in the phase plots. For $\omega=0.96471$ and
$\omega=0.97$ two solutions coexist, one stable (in blue, solid
line) and one unstable (in black, dashed curve). The upper panel
displays the time plot of the capsule motion for the two coexisting
solutions at $\omega=0.97$.}\label{fig-bif-diag-om}
\end{figure}

Let us begin our study with the investigation of the capsule system
without control. Specifically, we will carry out the numerical
continuation of the period-1 motion of the capsule model with
respect to the frequency of external excitation $\omega$ with $K=0$,
for the parameter range displayed in the boxed area of Fig.\
\ref{bifur_nocontrol}(a). The outcome of the continuation process is
shown in Fig.\ \ref{fig-bif-diag-om}. This figure presents the
periodic response of the uncontrolled capsule system as $\omega$ is
varied, with the vertical axis showing the contact time, i.e., the
time the oscillating mass is in contact with the secondary spring
$k_{2}$. As can be observed in the bifurcation diagram, small values
of excitation frequency $\omega$ produce stable period-1 solutions
(as depicted in the bottom left panel), represented by the solid
branch shown in the figure. By increasing $\omega$, a critical point
is found, corresponding to a period-doubling bifurcation (labeled
PD) of limit cycles. Here, the stability of the period-1 solution is
lost, giving rise to a family of stable period-2 solutions, as shown
in the bottom right panel of Fig.\ \ref{fig-bif-diag-om}. This panel
displays two coexisting solutions for $\omega=0.96471$, a stable
period-2 orbit (in solid blue) produced by the PD point and an
unstable period-1 solution (dashed black).

\begin{figure}[H]
\centering \psfrag{T}{\normalsize Contact
time}\psfrag{tau}{\normalsize$\tau_{d}$}
\psfrag{xr}{\normalsize$x_{r}$}\psfrag{vr}{\normalsize$v_{r}$}
\psfrag{a}{$\tau_{d}=0.52$}\psfrag{b}{$\tau_{d}=0.6$}\psfrag{c}{$\tau_{d}=2.1$}
\psfrag{t}{\normalsize$t$} \psfrag{xc}{\normalsize$x_{2}$}
\includegraphics[width=\textwidth]{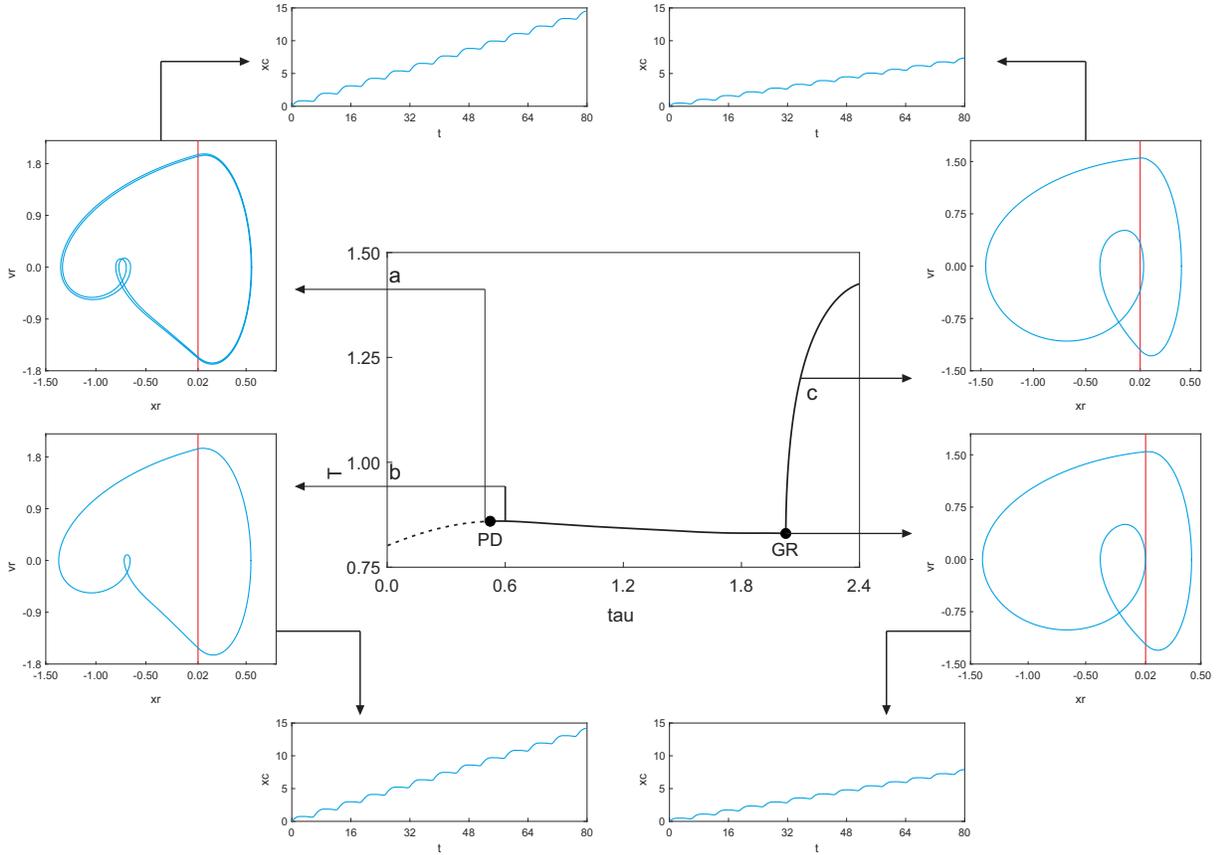}
\caption{Numerical continuation of periodic orbits of the capsule
system \eqref{eq-hybsys-dde-app} (considering time-delayed feedback
control) with respect to the control delay $\tau_{d}$, considering
the parameter values given in Fig.\ \ref{fig-bif-diag-om} (with
$K=0.5$ and $\omega=0.97$). The vertical axis presents the time the
oscillating mass is in contact with the secondary spring $k_{2}$.
The labels PD and GR mark period-doubling and grazing bifurcations
of limit cycles, located at $\tau_{d}=\tau^{\mbox{\tiny
GR}}_{d}\approx0.52329$ and $\tau_{d}\approx2.02531$, respectively.
The panels displayed on the sides show phase portraits and time
plots of the capsule displacement for selected values of the control
delay $\tau_{d}$.}\label{fig-bif-diag-tau}
\end{figure}

Let us suppose, for instance, that for practical reasons a higher
operation frequency is required, for example $\omega=0.97$, a value
not so far away from the period-doubling bifurcation detected above.
At this frequency, the system presents a chaotic behaviour as shown
in the center right panel depicted in Fig.\ \ref{fig-bif-diag-om},
possibly originating from a period-doubling cascade. As can be seen
in this phase plot, the stable chaotic response (in solid blue)
coexists with the original unstable period-1 solution (dashed
black). From a practical point of view, however, a period-1 behaviour
is in general more convenient, for instance due to a better
performance in terms of average speed of progression, as can be seen
in the time plot displayed in the upper panel in Fig.\
\ref{fig-bif-diag-om}, showing the capsule motion for the two
coexisting solutions at $\omega=0.97$. Therefore, one of the main
questions in the present is how to suitably employ the {\color{black}DFC} to stabilize this unstable period-1 response.
\begin{figure}[h!]
\centering \psfrag{xr}{\Large$x_{r}$}\psfrag{vr}{\Large$v_{r}$}
\psfrag{x1}{$\tau_{d}=0.52$}\psfrag{x2}{$\tau_{d}=\tau^{\mbox{\tiny
GR}}_{d}$}\psfrag{x3}{$\tau_{d}=2.1$}\psfrag{x4}{$\tau_{d}=2.8$}
\psfrag{x5}{$\tau_{d}=3.6$}\psfrag{x6}{$\tau_{d}=4.4$}
\includegraphics[width=\textwidth]{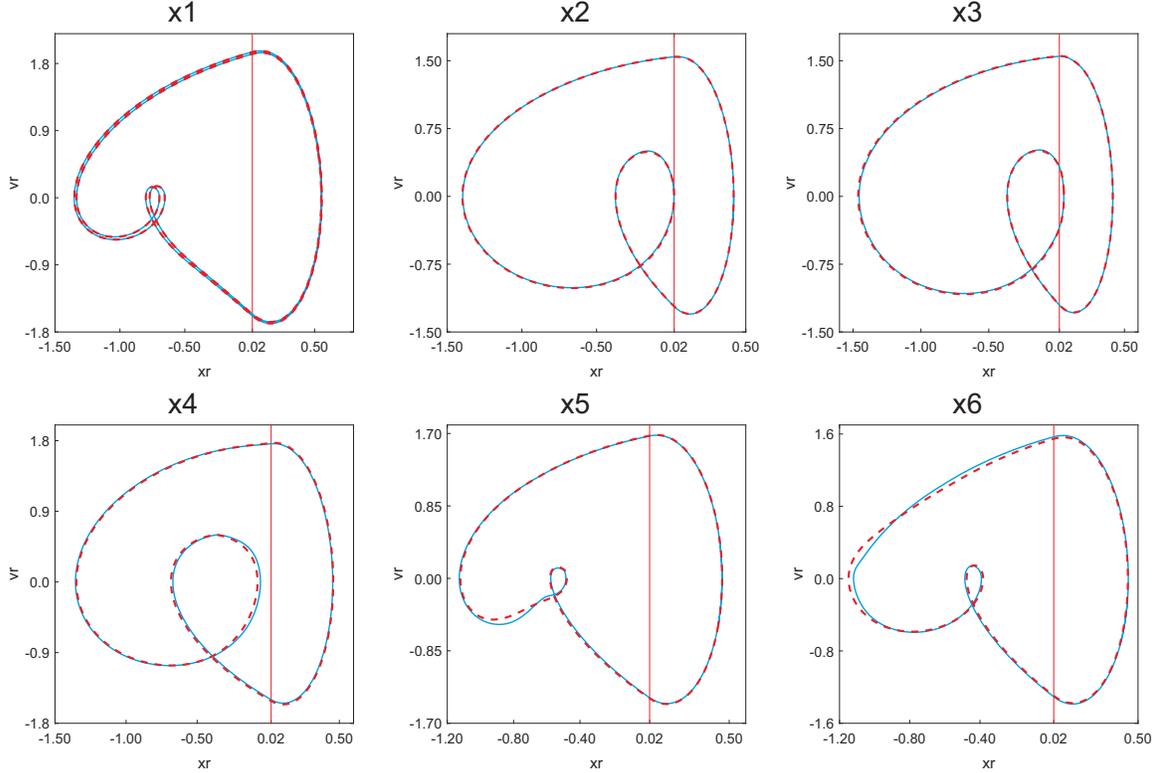}
\caption{{\color{black}Numerical comparison of the dynamical response of the
piecewise-smooth DDE \eqref{capsule_delaymodel} with the
approximating system of ODEs \eqref{eq-hybsys-dde-app}. The first
row shows the phase plots of the solutions obtained in Fig.\
\ref{fig-bif-diag-tau}. The lower row displays orbits obtained for
higher values of the delay $\tau_{d}$. The solutions to the original
system \eqref{capsule_delaymodel} and their approximations are
depicted in blue (solid line) and in red (dashed line),
respectively.}}\label{fig-delay-tests}
\end{figure}
We will now employ the numerical continuation platform COCO to
investigate the periodic response of the capsule system (with
the {\color{black}DFC}) using the approximating system
\eqref{eq-hybsys-dde-app}. Specifically, we will carry out the
numerical continuation of the unstable period-1 solution discussed
above with respect to the control delay $\tau_{d}$. The result is
shown in Fig.\ \ref{fig-bif-diag-tau}. As expected, a branch of
unstable period-1 solutions is found for low values of $\tau_{d}$.
{\color{black}With appearing of period-doubling bifurcation at $\tau_{d}\approx0.52329$, this solution becomes stable.}
{\color{black}As a consequence, the DFC successfully introduces the unstable desired period-1 response into the vibro-impact capsule system and ensures the system being stabilised on period-1 motion by inducing the period-doubling bifurcation.}
Upper and lower
left panels present phase plots computed at test points
$\tau_{d}=0.52$ and $\tau_{d}=0.6$ (before and after the
bifurcation), showing stable period-2 and period-1 orbits,
respectively. As the time delay increases, a grazing bifurcation
(GR) is found at $\tau_{d}\approx2.02531$, where a segment of the
periodic solution intersects tangentially the impact boundary
$x_{r}=\delta$. This grazing solution is plotted in the bottom right
panel shown in Fig.\ \ref{fig-bif-diag-tau}, while the upper right
panel shows a period-1 solution right after the grazing bifurcation
GR. According to these observations, the detected GR point defines a
boundary between two operation regimes: period-1 behaviour with one
impact with the secondary spring $k_{2}$ per excitation period and
period-1 solutions with two impacts per period. Therefore, our
numerical investigation reveals that the time-delayed feedback
control considered in this work has a direct effect on both the
stability the desired period-1 motion and impacting regimes in the
system.

{\color{black}
In order to test the reliability of the approximation scheme
\eqref{eq-hybsys-dde-app} employed for the numerical continuation
process described above, we will carry out a numerical comparison of
the dynamical behaviour of the original piecewise-smooth DDE
\eqref{capsule_delaymodel} with that of the approximating system of
ODEs \eqref{eq-hybsys-dde-app}, for different values of the delay
parameter $\tau_{d}$, see Fig.\ \ref{fig-delay-tests}. In this
diagram we present a series of orbits calculated from the original
piecewise-smooth DDE using the MATLAB solver `\texttt{dde23}' (solid
blue line) together with the approximating solutions obtained from
the discretization scheme \eqref{eq-hybsys-dde-app} (dashed red
line). For the range considered in the continuation study presented
in Fig.\ \ref{fig-bif-diag-tau} we can verify that the computed
solutions of the original DDE and those of the approximating ODE
system are very close to each other, for different values of
$\tau_{d}$, see first row of Fig.\ \ref{fig-delay-tests}. As the
time delay increases, however, it can be observed that the dynamical
response of the discretization scheme deviates from the solutions of
the original system, owing to the truncation error (i.e.\ the
$\bigo$-term) neglected from the Taylor expansion
\eqref{eq-Taylor-disc}. For this reason the numerical approach
proposed here is unable to provide reliable approximations for
arbitrary delays, and hence our study is restricted to small values
of $\tau_{d}$.}

\begin{figure}[H]
\centering
\psfrag{K}{\normalsize$K$}\psfrag{tau}{\normalsize$\tau_{d}$}
\psfrag{xr}{\normalsize$x_{r}$}\psfrag{vr}{\normalsize$v_{r}$}
\includegraphics[width=\textwidth]{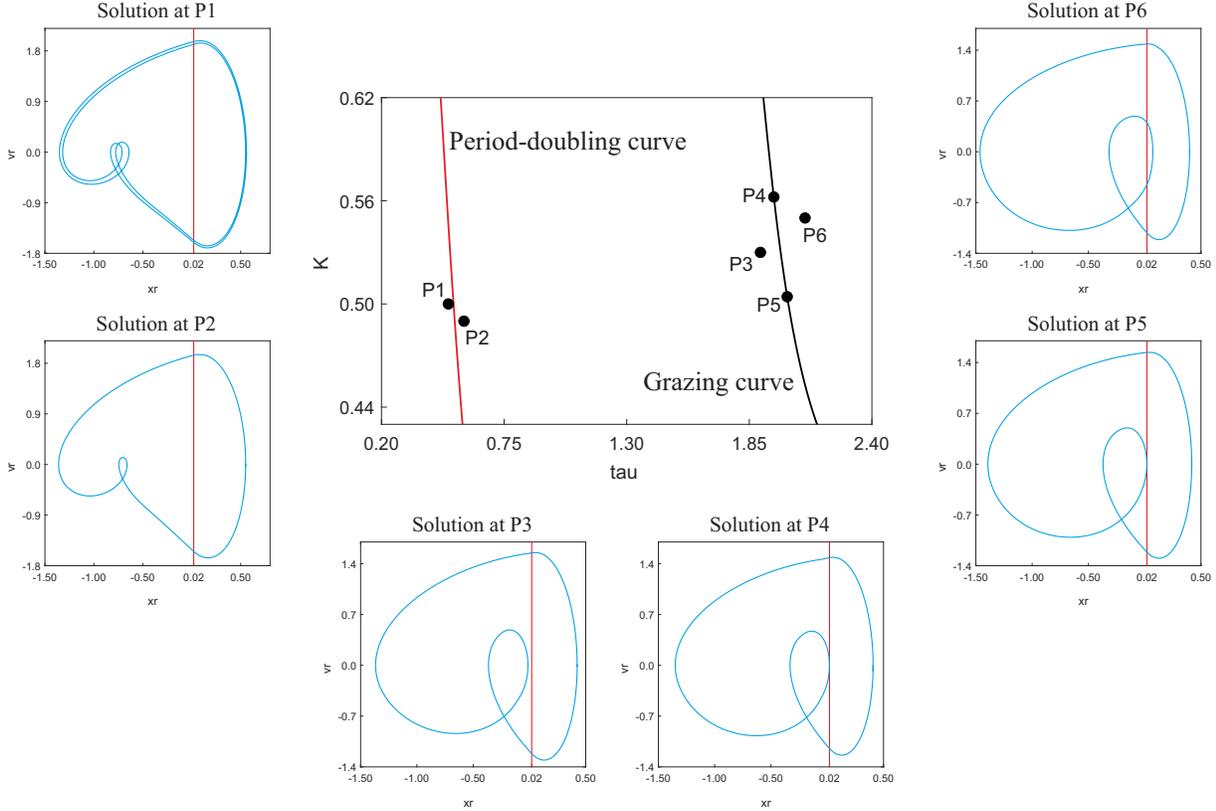}
\caption{ Two-parameter continuation of the period-doubling and
grazing bifurcations detected in Fig.\ \ref{fig-bif-diag-tau}, with
respect to the control parameters $\tau_{d}$ and $K$. Lateral and
lower panels display phase plots corresponding to the test points P1
($\tau_{d}=0.5$, $K=0.5$), P2 ($\tau_{d}=0.57$, $K=0.49$), P3
($\tau_{d}=1.9$, $K=0.53$), P4 ($\tau_{d}=1.96$, $K\approx0.56223$),
P5 ($\tau_{d}=2.02$, $K\approx0.50423$) and P6 ($\tau_{d}=2.1$,
$K=0.55$).}\label{fig-bif-diag-tau-k}
\end{figure}

Next, we will investigate the behaviour of the codimension-1
bifurcations discussed above when two control parameters are
perturbed, namely, the control delay $\tau_{d}$ and the control gain
$K$. In particular, a locus of grazing periodic orbits will be
computed by introducing an extra solution segment with a terminal
point satisfying the condition $\dot{x}_{r}=0$, where the relative
mass velocity becomes zero. In this way, a grazing solution can be
detected via the auxiliary boundary condition $x_{r}-\delta=0$, and
therefore a locus of such orbits can be obtained by freeing two
parameters during the continuation process, using the COCO-command
`coco\_xchg\_pars' \cite{dankowicz2013recipes}.

In Fig.\ \ref{fig-bif-diag-tau-k} we present the resulting red and
black curves corresponding to the two-parameter continuation of the
period-doubling and grazing bifurcations detected in Fig.\
\ref{fig-bif-diag-tau}, respectively. In this way, the
two-dimensional parameter space is locally divided into three
regions. The first region, located to left of the period-doubling
curve, gives combinations of $\tau_{d}$ and $K$ yielding unstable
period-1 response in the controlled system, and hence stable
solutions of higher period or even chaotic behaviour appear in the
system. A supercritical period-doubling bifurcation is produced upon
crossing this red line from the right, thereby creating stable
period-2 solutions, see for instance the one depicted in Fig.\
\ref{fig-bif-diag-tau-k} (upper left panel), computed for the test
point P1 ($\tau_{d}=0.5$, $K=0.5$). Between the period-doubling and
grazing curves a second region is defined, which corresponds to
parameter values producing a stable period-1 response with one
impact with the secondary spring $k_{2}$ per excitation period, as
can be seen in Fig.\ \ref{fig-bif-diag-tau-k} for the test points P2
($\tau_{d}=0.57$, $K=0.49$) and P3 ($\tau_{d}=1.9$, $K=0.53$).
Finally, a third region (to the right of the grazing curve) gives
$(\tau_{d},K)$ pairs producing stable period-1 solutions with two
impacts per period, see for instance the test orbit calculated for
P6 ($\tau_{d}=2.1$, $K=0.55$) which crosses the impact boundary
$x_{r}=\delta$ with $v_{r}>0$ twice. The phase plots computed for P4
($\tau_{d}=1.96$, $K\approx0.56223$) and P5 ($\tau_{d}=2.02$,
$K\approx0.50423$) are calculated on the critical boundary defined
by the grazing bifurcation curve (in black), showing solutions
making tangential contact with the discontinuity boundary
$x_{r}=\delta$.

\section{Conclusions}\label{sec-conclusion}
In this paper, {\color{black}the DFC with a constant delay} was
considered to control the multistability in a periodically forced
capsule system. The control aim was to improve the progression
efficiency of the capsule through switching the system to one of its
coexisting attractors with the highest progression speed. The
performance of the {\color{black}DFC} was studied numerically
by considering the energy consumption from the control signal and
the convergence time to the desired response. In addition, the complex dynamics of the controlled vibro-impact capsule system were exploited by using path-following
(continuation) methods for non-smooth delay differential
equations.

Our numerical studies revealed that the {\color{black}DFC}
with {\color{black}a constant delay} can reach the
period-1 response from the period-$3$ response for all the control
gains $K$ providing that they are sufficiently large. {\color{black}For a certain range of control gain,} the controlled system can converge to the desired motion rapidly
with only a relatively small energy expenditure. Thus, under this
control scenario, maintaining the control parameter $K$ at a lower
value in the feasible range may control the system at a desired
motion, and can also keep its energy expenditure at a lower level.
Our parameter studies also find that the {\color{black}DFC}
cam steer the system towards a period-1 motion for the delays
smaller than the excitation period with reasonable control inputs.
In addition, numerical continuation analysis for
non-smooth dynamical systems with time delays was employed to study
more detailed dynamics of the controlled capsule system
with a small time delay. When varying the main control parameters,
namely the time delay and the control gain, we observed
that the {\color{black}DFC} stabilised the period-1 motion in
a large region of these control parameters. In particular, the
feasible range of these parameters was identified between the
occurrences of a period-doubling and a grazing bifurcations via
two-parameter continuation.

\js{Because the studied system \eqref{capsule_delaymodel} with DFC \eqref{delu4capsule} is piecewise linear in all four regimes, trajectory segments are all known analytically such that one could in principle determine periodic orbits as solutions of exact finite systems of nonlinear algebraic systems of equations. However, determining the linear stability of these periodic orbits requires numerical discretisation of the orbit segments (in our case pseudo-spectral disretisation \eqref{eq-hybsys-dde-app} \cite{breda2016pseudospectral}), which we also use for bifurcation analysis. For all periodic orbits we detected and tracked, our numerical discretisation and event detection methods employ well-studied numerical methods fully within their regime of convergence, such that numerical discretization errors are small compared to the uncertainties fo our model. Consequently, all results regarding  existence, stability and bifurcations of periodic orbits and the transitions from attractors of the uncontrolled system to periodic orbits when applying DFC are robust with respect to the choice of numerical mmethods. Our analysis is unable to discover all coexisting attractors of the uncontrolled system, such that our approach cannot guarantee that DFC indeed makes the system globally monostable. However, analysis of basins of attraction as shown in Fig.~\ref{bistability} demonstrates that attractors with large basins are included in our study.}
{\color{black}The method and analyses discussed in the present work can be applied to study the dynamical systems with a similar structure as the vibro-impact capsule system} encountering time delay. The findings of this work can provide guidance for tuning the control parameters of the capsule system in experiment, e.g., \cite{guo2020self,liu2016experimental}, which could be the future work of this research project.

\section*{Acknowledgements}
This work has been supported by EPSRC under Grant No. EP/P023983/1.
Dr Zhi Zhang would like to acknowledge the financial support from
the University of Exeter for his Exeter International Excellence
Scholarship. Prof. Jan Sieber gratefully acknowledges support by
EPSRC Fellowship EP/N023544/1 and EPSRC grant EP/V04687X/1.

\section*{Compliance with ethical standards}

\section*{Conflict of interest}
The authors declare that they have no conflict of interest
concerning the publication of this manuscript.

\section*{Data accessibility}
The datasets generated and analysed during the current study are
available from the corresponding author on reasonable request.
\section*{Open access}
For the purpose of open access, the author has applied a CC BY
public copyright licence to any author accepted manuscript version
arising.

\bibliographystyle{ieeetr}
\bibliography{bibl}

\end{document}